\documentclass[letterpaper,twocolumn,10pt]{article}
\usepackage{usenix}

\usepackage[most]{tcolorbox}
\usepackage{graphicx} %
\usepackage{hyperref}
\usepackage{amsmath, amsthm}
\usepackage{amssymb}
\usepackage{nameref}
\usepackage[shortlabels]{enumitem}
\usepackage{todonotes}
\usepackage{graphicx}
\usepackage{float}
\usepackage{subfig}

\usepackage{mathtools}
\usepackage{xspace}
\usepackage{cleveref}
\usepackage{mdframed}
\usepackage{booktabs}   %
\usepackage{tabularx}
\usepackage{hyperref}
\usepackage{makecell}
\usepackage{setspace}
\usepackage{tikz}
\usepackage[edges]{forest}
\usepackage{makecell}

\newif \iffull \fulltrue
\newif \ifcomments \commentsfalse

\ifcomments
\newcommand{\andres}[1]{{\small\textsf{\color{green}{[Andrés: {#1}]}}}}
\newcommand{\ari}[1]{{\small\textsf{\color{red}{[Ari: {#1}]}}}}
\newcommand{\jay}[1]{{\small\textsf{\color{blue}{[Jay: {#1}]}}}}
\newcommand{\james}[1]{{\small\textsf{\color{olive}{[James: {#1}]}}}}
\newcommand{\sam}[1]{{\small\textsf{\color{purple}{[Sam: {#1}]}}}}
\newcommand{\ujval}[1]{{\small\textsf{\color{orange}{[ujval: {#1}]}}}}
\newcommand{\sarah}[1]{{\small\textsf{\color{teal}{[Sarah: {#1}]}}}}
\else
\newcommand{\andres}[1]{}
\newcommand{\ari}[1]{}
\newcommand{\jay}[1]{}
\newcommand{\james}[1]{}
\newcommand{\sam}[1]{}
\newcommand{\ujval}[1]{}
\newcommand{\sarah}[1]{}
\fi

\theoremstyle{definition}
\newtheorem{theorem}{Theorem}
\newtheorem{lemma}{Lemma} 
\newtheorem{definition}{Definition}

\newtheorem{remark}{Remark}
\theoremstyle{remark}

\newcommand{\ETH}{\texttt{ETH}}

\newcommand{\AMMfee}{\ensuremath{\mu}}

\hypersetup{
     colorlinks=true,      		%
     linkcolor=blue,        %
     citecolor=blue,     %
     filecolor=blue,      		%
     urlcolor=blue,           	%
}

\newcommand{\limitarrow}{\mathrel{\downarrow}}
\newcommand{\alg}{\textsf{ALG}}
\newcommand{\algSpace}{\mathbb{F}}
\newcommand{\algDist}{\textsf{ALG}_\textsf{dist}}
\newcommand{\algSample}{\textsf{ALG}_\textsf{sample}}
\newcommand{\leakage}{\mathcal{L}}
\newcommand{\players}{{\mathcal{P}}}
\newcommand{\player}{\ensuremath{P}}
\newcommand{\advA}{\mathcal{A}}
\newcommand{\stateSpace}{\mathbb{S}}
\newcommand{\bountysystem}{\mathcal{B}}
\newcommand{\stateSpaceSubset}{\tilde{\mathbb{S}}}
\newcommand{\state}{\ensuremath{\mathcal{S}}}
\newcommand{\balance}{{\textsf{balance}}}
\newcommand{\action}{{\textsf{action}}}
\newcommand{\transition}{{\textsf{transition}}}
\newcommand{\price}{{\textsf{price}}}
\newcommand{\trade}{\ensuremath{\tau}}
\newcommand{\tradeSpace}{\mathbb{T}}
\newcommand{\block}{B}

\newcommand{\assetSpace}{\mathbb{A}}
\newcommand{\asset}{A}
\newcommand{\DAO}{\ensuremath{\mathcal{C}}\xspace}

\newcommand{\transcript}{\mathcal{T}}

\newcommand{\getsr}{{\:{\leftarrow{\hspace*{-3pt}\raisebox{.75pt}{$\scriptscriptstyle\$$}}}\:}}
\newcommand{\sysnamelongu}{\underline{Co}llective \underline{In}vestment \underline{Alg}orithms\xspace}

\newcommand{\tradeoffName}{CoinAlg Bind\xspace}
\newcommand{\sysnameshort}{CoinAlgs\xspace}
\newcommand{\sysnameshortsingle}{CoinAlg\xspace}

\newcommand{\adv}{{\ensuremath{\cal A}}}
\newcommand{\balAMM}
{\textsf{bal}^{\textsf{AMM}}}
\newcommand{\balD}{\textsf{bal}^{\DAO}}
\newcommand{\balAdv}{\textsf{bal}^{\adv}}
\newcommand{\pairbal}[2]{\big[#1 \mbox{ USD}, #2 \mbox{ TOK}\big]}
\newcommand{\buy}[1]{{\ensuremath{\textsf{buy}}(#1)}}
\newcommand{\plim}
{{\ensuremath{p_\textsf{lim}}}} 
\newcommand{\buyl}[2]{{\ensuremath{\textsf{buy}}(#1)_{\limitarrow #2}}}
\newcommand{\sggame}{{\ensuremath{\textsf{Sdwch}}}\xspace}
\newcommand{\ultgame}{{\ensuremath{\textsf{U-Sdwch}}}\xspace}

\newcommand{\oraclehat}{\ensuremath{{\hat{\mathcal{O}}}}}
\newcommand{\oracle}{\ensuremath{\mathcal{O}}}
\newcommand{\epsilonMin}{\epsilon_{\textsf{min}}}
\newcommand{\concat}{\: \| \:}

\newcommand{\distinguisher}{\textsf{Dist}}
\newcommand{\fullPrivFcnLabel}{\textsf{priv}}

\newcommand{\assetFcnLabel}{\textsf{asset}}
\newcommand{\directionFcnLabel}{\textsf{dir}}
\newcommand{\fullTransparencyFcnLabel}{\textsf{transp}}
\newcommand{\discount}{\delta}

\newcommand{\procedurev}[1]{\underline{{#1}:}\smallskip}

\newcommand{\gamesfontsize}{\small}
\newcommand{\stretchval}{1.2}

\newcommand{\fpage}[2]{\framebox{\begin{minipage}{#1\textwidth}\setstretch{\stretchval}\gamesfontsize #2 \end{minipage}}}

\newcommand{\FairnessGame}{\textsc{Fair}}

\newcommand{\FairnessGameShort}{\textup{fair}}
\newcommand{\Advantage}[2]{\textnormal{Adv}^{#1}_{#2}}
\newcommand{\Prob}[1]{\Pr\left[\: #1 \:\right]}

\newcommand{\creturn}{\mathbf{return\;}}
\newcommand{\cfor}{\mathbf{for\;}}

\newcommand{\ind}{\hspace*{1.5em}}

\makeatletter
\renewcommand{\paragraph}{%
  \@startsection{paragraph}{4}%
  {\z@}{1.2ex \@plus 1ex \@minus .2ex}{-1em}%
  {\normalfont\normalsize\bfseries}%
}
\makeatother

\definecolor{keyInsightColor}{HTML}{DDCC77}
\newtheorem{keyInsightEnv}{Key Insight}
\newenvironment{keyInsight}
  {\begin{tcolorbox}[colframe=keyInsightColor]
   \begin{keyInsightEnv}}
  {\end{keyInsightEnv}
   \end{tcolorbox}}

\definecolor{yesColor}{HTML}{117733}
\definecolor{noColor}{HTML}{882255}

\newlength{\saveparindent}
\newlength{\saveparskip}
\newcounter{ctr}

\newenvironment{newenum}{%
\begin{list}{{\rm (\arabic{ctr})}\hfill}{\usecounter{ctr} \labelwidth=17pt%
\labelsep=5pt \leftmargin=22pt \topsep=3pt%
\setlength{\listparindent}{\saveparindent}%
\setlength{\parsep}{\saveparskip}%
\setlength{\itemsep}{2pt} }}{\end{list}}

\title{The \tradeoffName: \\Profitability-Fairness Tradeoffs in Collective Investment Algorithms}

\author{
{\rm Andrés Fábrega$^*$}\\
Cornell Tech, IC3
\and
{\rm James Austgen$^*$}\\
Cornell Tech, IC3
\and
{\rm Samuel Breckenridge}\\
Cornell Tech, IC3
\and
{\rm Jay Yu}\\
Pantera Capital 
\and
{\rm Amy Zhao}\\
Ava Labs, IC3
\and
{\rm Sarah Allen}\\
Flashbots, IC3
\and
{\rm Aditya Saraf}\\
Cornell Tech, IC3
\and
{\rm Ari Juels}\\
Cornell Tech, IC3
}
\date{}

\begin{document}

\maketitle

\def\thefootnote{*}\footnotetext{These authors contributed equally to this work.}
\renewcommand*{\thefootnote}{\arabic{footnote}}

\pagestyle{plain}

\begin{abstract}
\textit{\sysnamelongu} (\sysnameshort) are increasingly popular systems that deploy shared trading strategies for investor communities. Their goal is to democratize sophisticated---often AI-based---investing tools. 

We identify and demonstrate a fundamental \textit{profitability-fairness tradeoff} in \sysnameshort that we call the \textit{\tradeoffName}: \sysnameshort cannot ensure  economic fairness without losing profit to arbitrage. 

We present a formal model of \sysnameshort, with definitions of \textit{privacy} (incomplete algorithm disclosure) and \textit{economic fairness} (value extraction by an adversarial insider). We prove two complementary results that together demonstrate the \tradeoffName. First, \textit{privacy} in a \sysnameshortsingle is a precondition for insider attacks on economic fairness. Conversely, in a game-theoretic model, lack of privacy, i.e., \textit{transparency}, enables arbitrageurs to erode the profitability of a \sysnameshortsingle. 

Using data from Uniswap, a decentralized exchange, we empirically study both sides of the \tradeoffName. We quantify the impact of arbitrage against transparent \sysnameshort. We  show the risks posed by a private \sysnameshortsingle: Even low-bandwidth covert-channel information leakage enables unfair value extraction.

\end{abstract}

\section{Introduction}
\label{sec:intro}

This paper initiates the study of tradeoffs in what we call \textit{\sysnamelongu} (\textit{\sysnameshort})---algorithms shared by a community of users to dictate or execute financial transactions on their behalf in pursuit of profitable investment strategies. 

\sysnameshort have long existed in traditional finance. For example, quantitative strategies at firms like BlackRock Systematic or Renaissance Technologies execute trades across many client portfolios. Variants also exist in the form of robo-advisers, e.g., Wealthfront\iffull{, that use shared AI engines for financial planning.}\fi

\sysnameshort have recently gained traction outside traditional financial firms.  
AI-based trading software packages for individual investors have become available for both crypto and traditional financial assets, e.g.,~\cite{pragmaticcoders2025aitools,werner2025aibot}. 
In decentralized finance (DeFi), ElizaOS (token AI16Z), ``one of the first known AI-controlled venture capital funds,'' achieved a peak market capitalization of \$2+ billion~\cite{crypto_elizaos,coinbase_ai16z}, while Truth Terminal, an AI agent that drives community trading with tweets on X, initiated the launch of Goatseus Maximus (GOAT), a memecoin with a peak market capitalization of \$1+ billion~\cite{coindesk_truth_terminal_2024,truth_terminal_x}. 

\sysnameshort in general promise to place institutional-grade investment strategies or specialized trading intelligence within the reach of retail investors. Newly emerging ones, though---\iffull{direct-to-}\fi{}consumer trading tools and AI-based crypto platforms---lack the investor protections of regulated intermediaries. We therefore ask: What design tradeoffs and risks are inherent in \sysnameshort?

In this work, we prove the existence of a fundamental dilemma for \sysnameshort' designers and users. A {\sysnameshortsingle}’s trading strategy must be at least partially \textit{private} or else \textit{transparent}. Both options, though, place investor profits at risk: 

\begin{itemize}
    \item If \textbf{private}, a \sysnameshortsingle can leak information to insiders (through, e.g., a covert channel), enabling unfair and potentially undetectable insider \textit{value extraction} from investors.
    \item If \textbf{transparent}, a \sysnameshortsingle leaks information publicly, exposing it to arbitrage and degraded profits---what we term a \textit{cost of transparency}.  
\end{itemize}

We call this tension the \textit{\tradeoffName}: Any \sysnameshortsingle must sacrifice either economic fairness or profitability.

\subsection{Our Results}

Our work represents the first formal treatment of the tradeoff expressed by the \tradeoffName. 
It comprises three main results, on the privacy, transparency, and empirical investigation of \sysnameshort.

\paragraph{\sysnameshortsingle privacy and  economic (un)fairness.} 
We define a \sysnameshortsingle as \textit{private} if there is a non-negligible statistical distance  between a probability distribution over expected trades induced by a public view of \sysnameshortsingle and the actual, executed trades of the \sysnameshortsingle. In other words, the public view is insufficient to predict its transactions. We present a formal definition of \sysnameshortsingle privacy in our work. 

We also present a definition of \textit{fairness} as a measure of \textit{value extraction} by an adversary that uses inside knowledge to exploit (e.g., sandwich) the transactions of a \sysnameshortsingle. This definition is an insider variant of similar concepts in decentralized finance (e.g., ``miner / maximal extractable value'' (MEV))~\cite{babel2023clockwork,daian2020flash,qin2022quantifying,zhou2023sok}.

Our main result here (\Cref{thm:unfair_implies_private}) is that \textit{economic unfairness implies privacy}. Intuitively, privacy is necessary to give an advantage to an adversarial insider. (The converse doesn't always hold: Privacy need not mean that an adversary has actionable insider information.)

\paragraph{Transparency in \sysnameshort.} We also explore \sysnameshort that are not private, but \textit{transparent}. Public access to a transparent \sysnameshortsingle---e.g., open-sourcing its algorithm---enables advance knowledge of its transactions. We demonstrate that this exposes a \sysnameshortsingle to arbitrage by an adversary $\adv$. Our transparency model in practice relies only on the weak assumption that $\adv$ can simulate the \sysnameshortsingle; it doesn't require $\adv$ to reverse-engineer the \sysnameshortsingle.

Modeling the interaction between $\adv$ and a transparent \sysnameshortsingle as an ultimatum game, we show that $\adv$ can use a \textit{grim-trigger} strategy---the threat of invalidating the {\sysnameshortsingle}'s transactions---to force \sysnameshortsingle exposure to arbitrage by $\adv$ (Theorem~\ref{thm:ult-game-eq}). The ideas behind our model may be of independent interest, as they involve a \emph{repeated game} where \emph{intentional transaction invalidation} forms the basis of a trigger strategy.

\paragraph{Experiments.}

Using historical transactions in a decentralized exchange (Uniswap V3) on Ethereum for a period of over one year (June 2024 to July 2025), we simulate a profitable \sysnameshortsingle that correctly predicts future asset prices. We study both sides of the \tradeoffName. We demonstrate that even minimal information leakage from a private \sysnameshortsingle enables significant value extraction. We show conversely that even moderate defensive strategies against arbitrage incur a significant cost of transparency and thus eroded profits.

\bigskip

While we focus on crypto \sysnameshort for concreteness, our basic results apply more generally. For example, they 
emphasize why the investor protections of traditional finance are important to protect against value-extraction attacks in private \sysnameshort.

\subsection{Contributions}

In summary, our work initiates the study of the profitability-fairness tradeoff in \sysnameshort---what we term the \tradeoffName. After presenting background on \sysnameshort (\Cref{sec:what-are-coinalgs}) and the current landscape of their use, we make the following contributions:

\begin{itemize}
    \item \textbf{Privacy and fairness:} Based on a general formal model of \sysnameshort (\Cref{sec:definition}), we define \textit{privacy} and \textit{fairness} in \sysnameshort and prove that privacy is a precondition for unfairness and thus insider value extraction (\Cref{sec:privacy}).
    \item \textbf{Transparency:} Through game-theoretic modeling of a \sysnameshortsingle and an adversarial arbitrageur, we establish that \textit{lack} of privacy, i.e., \textit{transparency}, degrades \sysnameshortsingle profitability, imposing a \textit{cost of transparency} (\Cref{sec:transparency}).
    \item \textbf{Experiments:} We present experiments (\Cref{sec:empirical_transparency}) based on historical Ethereum markets that offer empirical support for our analytic results about privacy and fairness and the cost of transparency.
    \item \textbf{Guardrails:} Recognizing that despite their shortcomings, \sysnameshort are an inevitable part of the investing landscape, we briefly propose two practical guardrails that may offer \textit{heuristic} protection for \sysnameshortsingle investors against value extraction in private \sysnameshort (\Cref{sec:guardrails}).  
\end{itemize}

Our review of related work (\Cref{sec:related}) underscores the lack of existing treatment of the \tradeoffName, as well as the novelty of our game-theoretic model in the DeFi setting. We conclude (\Cref{sec:conclusion}) that the \tradeoffName expresses an intrinsic limitation in the design of collective investment algorithms: a choice is necessary between the fairness risks of privacy and the exposure to arbitrage of transparency. 

\section{What are CoinAlgs?}
\label{sec:what-are-coinalgs}
\noindent Two primary characteristics define \sysnameshort. They are:
\begin{enumerate}
    \item \emph{Collective}: Users in a \sysnameshortsingle are bound by a \emph{shared} trading logic that drives investments.
    \item \emph{Algorithmic}: The trading logic of a \sysnameshortsingle is \emph{algorithmic} in nature.
\end{enumerate}
In other words, \sysnameshort are \emph{algorithms that drive collective investment actions}. These two defining properties are what give relevance to the \tradeoffName: collectivity introduces fiduciary and thus fairness obligations, while the algorithmic nature introduces the potential for transparency or privacy.

\paragraph{Categories of \sysnameshort.} A number of seemingly unrelated financial services---across both traditional finance and Web3---come under the umbrella of \sysnameshort. Key among these are:

\smallskip
\noindent\emph{Traditional-finance algorithmic trading.} Collective management of investor assets is a well-established practice in traditional finance (a.k.a~TradFi), including at high-frequency trading (HFT) firms, hedge funds, and quantitative investment firms. Their platforms are often explicitly~\sysnameshort, pooling investors' assets and executing trades algorithmically on their behalf. 

Other examples of TradFi \sysnameshort are more subtle, involving investors sharing trading algorithms but acting independently. Robo-advisors---offered by many established TradFi firms (Vanguard, Schwab, etc.)---are an example, although customized to individual goals. eToro, a popular social-trading platform, allows explicit investor-to-investor replication of trading algorithms.

\smallskip
\noindent\emph{Direct-to-investor platforms.} 
Direct-to-investor algorithmic-trading platforms---such as Composer and QuantConnect for equities and Zignaly for crypto---enable individual investors to develop, share, or license trading algorithms while retaining independent control of their assets. As they allow communities of investors to share trading logic, they constitute \sysnameshort. 
\smallskip

\smallskip
\noindent \emph{AI-powered investment DAOs and communities.} In DeFi, there has been a flurry of interest in coupling AI-powered investment algorithms with Decentralized Autonomous Organizations (DAOs)---smart contracts in which investors pool funds. Prominent examples of such \sysnameshort include AI16Z and AI-XBT. A less obvious form of \sysnameshortsingle is Terminal of Truths, described in~\Cref{sec:intro}. 

\bigskip
Among these three types of \sysnameshort, those in
TradFi are \textit{custodial}, meaning that firms hold investors' funds as broker-dealers and are subject to investor-protection regulations. Direct-to-investor and DeFi \sysnameshort are typically \textit{non-custodial}, placing them outside traditional investor-protection regimes. 

Our definition of \sysnameshort excludes many popular financial services. For example, index funds in TradFi are not \sysnameshort: they do not operate algorithmically, so there is no opportunity for transparency.

\paragraph{Properties of \sysnameshort.} As our examples illustrate, \sysnameshort  can differ along two dimensions.

First, the way in which \emph{user assets are managed}. This can be \emph{consolidated}, which means that assets are pooled together in a central location (e.g., an AI-powered DAO, quantitative investment firm, etc.); or it can be \emph{fragmented}, which means that assets are traded individually (e.g., robo advisors). 

The second dimension is the way in which the \sysnameshortsingle's \emph{trading algorithm is accessed}. This can once again be consolidated or fragmented: the former refers to when the algorithm is in a single, central location (e.g., robo-advisors); and the latter to when multiple copies are distributed (e.g., open-source trading bots).

\begin{figure}[t!]
    \centering
\tikzset{every picture/.style={line width=0.75pt}} %

\begin{tikzpicture}[x=0.75pt,y=0.75pt,yscale=-1,xscale=1]

\draw  [line width=2.25]  (159,116) -- (229,116) -- (229,186) ;
\draw  [line width=2.25]  (299,116) -- (229,116) -- (229,46) ;

\draw (301,108) node [anchor=north west][inner sep=0.75pt]  [font=\footnotesize] [align=left] {\begin{minipage}[lt]{51.4pt}\setlength\topsep{0pt}
\begin{center}
{\fontfamily{ptm}\selectfont {\color[HTML]{CC6677} \scriptsize Consolidated (assets)}}
\end{center}

\end{minipage}};
\draw (88,108) node [anchor=north west][inner sep=0.75pt]  [font=\footnotesize] [align=left] {\begin{minipage}[lt]{48.54pt}\setlength\topsep{0pt}
\begin{center}
{\color[HTML]{CC6677} \scriptsize {\fontfamily{ptm}\selectfont Fragmented (assets)}}
\end{center}

\end{minipage}};
\draw (187,22) node [anchor=north west][inner sep=0.75pt]  [font=\footnotesize] [align=left] {\begin{minipage}[lt]{60.28pt}\setlength\topsep{0pt}
\begin{center}
{\fontfamily{ptm}\selectfont {\color[HTML]{CC6677} \scriptsize Consolidated (algorithm)}}
\end{center}

\end{minipage}};
\draw (188,188) node [anchor=north west][inner sep=0.75pt]  [font=\footnotesize] [align=left] {\begin{minipage}[lt]{57.41pt}\setlength\topsep{0pt}
\begin{center}
{\fontfamily{ptm}\selectfont {\color[HTML]{CC6677} \scriptsize Fragmented (algorithm)}}
\end{center}

\end{minipage}};
\draw (157,79) node [anchor=north west][inner sep=0.75pt]  [font=\footnotesize] [align=left] {{\fontfamily{ptm}\selectfont {\footnotesize Robo-advisors}}};
\draw (135,59) node [anchor=north west][inner sep=0.75pt]  [font=\footnotesize] [align=left] {{\fontfamily{ptm}\selectfont {\footnotesize AI gambling agents}}};
\draw (235,89) node [anchor=north west][inner sep=0.75pt]  [font=\footnotesize] [align=left] {{\fontfamily{ptm}\selectfont {\footnotesize AI-powered DAOs}}};
\draw (140,140) node [anchor=north west][inner sep=0.75pt]  [font=\footnotesize] [align=left] {{\fontfamily{ptm}\selectfont {\footnotesize Open-source bots}}};
\draw (245,65) node [anchor=north west][inner sep=0.75pt]  [font=\footnotesize] [align=left] {{\fontfamily{ptm}\selectfont {\footnotesize Quant. trading firms}}};

\end{tikzpicture}
\caption{Example \sysnameshort and whether user assets and their trading algorithm are consolidated or fragmented.}\label{fig:coinalgs-examples}
\end{figure}
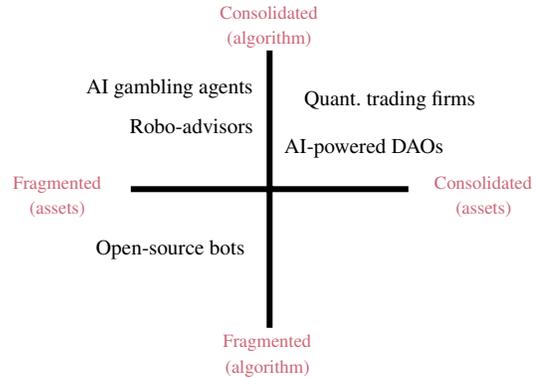

We show a few example \sysnameshort in Figure~\ref{fig:coinalgs-examples}, categorized by their type of asset management and algorithm. Note that the bottom-right quadrant is empty; we are not aware of any \sysnameshort that consolidate assets but have a fragmented algorithm.

\subsection{\sysnameshort in the Wild}
We observe three patterns in the behavior of \sysnameshort in practice that inform our investigation.

\begin{figure}[t!]
    \centering
    \begin{tabular}{l|rrl}
     \toprule
     \textbf{Robo-advisor} & \textbf{AUM} & \textbf{No. clients} & \textbf{Visibility} \\ 
     \midrule
    Vanguard & \$206.6B & 1.1M & Private \\
    Schwab & \$65.8B & 262K & Private \\
    Betterment & \$26.8B & 615K & Private \\
    Wealthfront & \$21.4B & 307K & Private \\
    Acorns & \$4.7B & 4.4M & Private \\
    \bottomrule
\end{tabular}
\caption{Overview of popular robo-advisors, showing their assets under management (AUM), number of individual clients, and whether they are private or transparent.
\iffull
Numbers for AUM and number of clients are as of an August 2024 report by Forbes~\cite{robo-advisors-stats}.
\else
\fi}
\label{fig:roboadvisors-takeaways-stats}
\end{figure}

\begin{figure*}[t!]
    \centering
    \begin{tabular}{l|rrll}
     \toprule
     \textbf{AI DAO} & \textbf{Peak MC} & \textbf{Peak AUM} & \textbf{Privacy} & \textbf{Major Holdings}\\ 
     \midrule
    ai16z / Eliza & \$2.7B & \$22.9M & Private & ai16z, degenai, FXN \\
    Aiccelerate DAO & \$274.9M & \$171K & Private & SOL, AICC, Ropirito, bAInance \\
    Big PharmAI & \$57.9M & \$694.8K & Private & BIO, SOL, PYTHIA, OMIPAL\\
    AI XBT & \$4.7B & \$755.4M & Private & VIRTUAL, aixbt, ETH, CLANKER, USDC, LUNA\\
    Vader AI & \$142.6M & \$27.4K & Private & VIRTUAL, USDT, WETH \\
    Axelrod & \$22.6M & \$2.2K & Private & ETH, VIRTUAL, AXR, doginme \\
    Terminal of Truths & \$1.3B & \$26.8M & Private & Fartcoin, GOAT, Shegen, Mana, Zerebro\\
    \bottomrule
\end{tabular}
\caption{Overview of AI-powered DAOs, showing their peak market cap (MC), peak assets under management (AUM), whether they are private or transparent, and their major asset holdings.}
\label{fig:ai-daos-takeaways-stats}
\end{figure*}

\paragraph{Pattern \#1: Significant interest.} \sysnameshort have been the source of much financial attention and interest. For example, in traditional finance, robo-advisors have become a prominent industry, being offered by leading financial services companies (e.g., Vanguard, Wealthfront, Charles Schwab, etc.) and managing large amounts of 
\iffull
assets: as of an August 2024 report by Forbes, the top 10 robo-advisors by AUM manage over \$350 billion and have over 6.7 million individual clients~\cite{robo-advisors-stats}. See Figure~\ref{fig:roboadvisors-takeaways-stats} for an overview of a few popular robo-advisors, all within the top 10 by AUM.
\fi

DeFi \sysnameshort---in particular, AI-powered DAOs---have also seen significant interest, skyrocketing in popularity in early 2025, when they began to feature prominently in Web3 discourse. Unlike robo-advisors, however, there are presently no landscape surveys of AI-powered DAOs reflecting the scale of this industry. Therefore, we performed a small-scale measurement study to understand the popularity of \sysnameshort. We display a snapshot of our results in Figure~\ref{fig:ai-daos-takeaways-stats}, which show that, like in traditional finance, significant capital is tied to Web3 \sysnameshort.

\paragraph{Pattern \#2: Privacy.} Most \sysnameshort operate with proprietary algorithms that are not publicly disclosed. As we show in Figure~\ref{fig:roboadvisors-takeaways-stats}, all existing robo-advisors are private, since users interact with them through an online interface. Similarly, as shown in Figure~\ref{fig:ai-daos-takeaways-stats}, AI-powered DAOs are also mostly private. 
\iffull 
Platforms such as daos.fun provide data on token valuation, portfolio composition, and trading history, but offer no transparency into underlying decision logic.
\fi

\paragraph{Pattern \#3: Volatile investments in DeFi.} DeFi \sysnameshortsingle portfolios skew heavily towards speculative assets, with a preference for emerging tokens. 
As shown in Figure~\ref{fig:ai-daos-takeaways-stats}, many AI-powered DAOs have at least one ``memecoin'' in their main portfolio holdings, motivating study of both stable and volatile assets in \sysnameshort.

\smallskip
\paragraph{\sysnameshort in TradFi vs.~DeFi} While \sysnameshort exist in both TradFi and DeFi, we focus on the latter in this work for two reasons. First, DeFi \sysnameshort lend themselves to \emph{crisper modeling}---blockchains provide natural abstractions which are much simpler than those of traditional markets---and are more amenable to formal study. Second, the open nature of DeFi provides a natural testbed for \emph{empirical studies} of \sysnameshort (see, e.g.,  Section~\ref{sec:empirical_transparency}), due to better data availability supporting more realistic experiments and simulations.

We stress, however, that our results point to a fundamental tension in \sysnameshort that applies broadly, including to TradFi. In TradFi, the \tradeoffName is addressed via legally required \emph{investor protections}, which allow \sysnameshort to be opaque in order to safeguard profits. Non-custodial \sysnameshort raise strong concerns because they lack these protections.

\section{Formal Model for \sysnameshort}\label{sec:definition}
To study the \tradeoffName theoretically, we first present a formal model for \sysnameshort more broadly. 
\iffull
We introduce a general model for blockchain systems (drawing inspiration from prior work~\cite{babel2023clockwork,heimbach2022eliminating}), followed by our definition for \sysnameshort in the context of this model, including a formal definition of privacy for \sysnameshort.
\else
\fi

\paragraph{System abstraction.} We denote the \emph{state} of the system (blockchain) by $\state$, and the universe of all possible states by $\stateSpace$. 
\iffull
Valid \emph{state transitions} are defined by the function $\transition : \stateSpace \times \stateSpace \to \{0, 1\}$, which returns 1 if and only if the second input state is reachable from the first. We call a finite sequence of states $\transcript = (\state_1, ..., \state_s)$ a \emph{transcript}, and say that a transcript is \emph{valid} if $\transition(\state_i, \state_{i+1}) = 1$ for every $i \in [q)$.

The system is composed of a set of \emph{players} $\players = \{\player_1, ..., \player_p\}$, which hold positions on a set of tradable assets $\assetSpace = \{\asset_1, ..., \asset_a\}$. A particular system state is thus defined by the balance of all players across all assets.\footnote{Technically, system state is also defined by storage data associated with players, but this is not relevant to our results, and so we omit it for clarity.} We define the function $\balance : \players \times \stateSpace \times \assetSpace \to \mathbb{N}$,\footnote{We follow the convention that $0 \in \mathbb{N}$ throughout the paper.} which maps a player to their holdings of an asset at a given state. 
\fi
We denote trades via tuples of the form $\trade \in \tradeSpace \coloneqq \player \times \stateSpace \times \assetSpace^2 \times \mathbb{N}$; for example $(\player, \state, \asset_1, \asset_2, N)$ indicates a swap of $N$ units of $\asset_1$ for $\asset_2$ by player $\player$ at state $\state$.
The state transition as a result of trade $\trade$ is denoted by the function $\action : \tradeSpace \times \stateSpace \to \stateSpace \cup \{\bot\}$, which returns $\bot$ if the trade is not valid in the current state (e.g., the player does not have enough units of the asset being sold).
\iffull
The number of units acquired during the trade is a function of current \emph{asset prices}, which we model via function $\price : \assetSpace \times \stateSpace \to \mathbb{R}_{>0}$, which returns the cost of one unit of the asset in terms of the system's primary asset (e.g., ETH in Ethereum).
\else
\fi

State transitions are triggered at fixed intervals (e.g., approximately 12 seconds for Ethereum), which execute (a subset of) all trades accumulated after the previous state transition. Given the present state $\state_0$ and a \emph{block}, i.e., an ordered sequence of trades $\block = (\trade_1, ..., \trade_q)$, the resulting system state is determined by $\action(\trade_q, \state_{q-1})$, where $\state_{i} = \action(\trade_{i}, \state_{i-1})$; we overload notation and use the shorthand $\action(\block, \state)$ to denote this process. Note that this returns $\bot$ if any of the underlying calls to $\action$ return 
\iffull 
$\bot$, in which case we say that the block is not valid. 
\fi
The trades that are included in a block, and their order, are determined by a system-specific protocol.

\paragraph{\sysnameshort model.} In our system abstraction, a \sysnameshortsingle $\DAO$ is a player in the system that makes trades according to an \emph{investment strategy} $\alg : \stateSpace \to \tradeSpace \cup \{\bot\}$, which is an algorithm that returns a trade to be made for a given input state (or $\bot$, if no trades would be made). We denote the space of all possible investment strategy functions by $\algSpace$. Critically, the behavior of $\DAO$ is fully determined by $\alg$, i.e., $\DAO$ will always make a trade output by $\alg(\state)$ in state $\state$. 

The trading algorithm $\alg$ may be deterministic or randomized. To capture this, we model $\alg$ as consisting of two logical subroutines: (1) $\algDist(\state)$, which (deterministically) outputs a probability distribution $\pi_\state$ over $\tradeSpace$; followed by (2) $\algSample(\pi_\state)$, which samples a specific trade $\trade$ from probability distribution $\pi_\state$. So, for example, a \emph{deterministic} \sysnameshortsingle is one for which $\algDist(\state)$ always outputs a degenerate distribution.

\paragraph{Privacy of  \sysnameshort.} We now formally define privacy for \sysnameshort, a necessary precondition for the principled analysis of the \tradeoffName.

At a high level, we characterize the privacy of $\DAO$ in terms of \emph{the ability of players to predict $\DAO$'s trades in advance}. Notice that our privacy notion is framed in terms of \emph{capabilities} of players (i.e., predict trades) as opposed to specific \emph{properties} of $\DAO$ (e.g., whether $\DAO$'s code is open-source or not). As we show in future sections, predicting trades is at the core of the \tradeoffName. We thus focus directly on the impact of (lack of) privacy, while abstracting away the mechanisms of how information is disclosed. 

For instance, say that $\DAO$ is fully transparent, so players can perfectly predict (distributions of) trades in advance. This could stem from the specification of $\alg$ being open-source (i.e., players having white-box access to $\alg$), or from players having just black-box, but unfettered, query access to $\alg$. For both, what matters is that players can accurately predict
\iffull 
trades (which, as we show in Section~\ref{sec:transparency}, undermines $\DAO$'s profits)
\else 
\fi 
trades, and not the mechanics of \emph{how} this information is obtained.

To model this formally, we introduce the notion of \emph{public functions}, which ``flatten'' $\alg$ to capture properties of future trades that are revealed to the public. (Without loss of generality, we assume, for now, that all players share the same knowledge of $\alg$.\footnote{For example, we can define the global knowledge of $\alg$ as the union of the individual knowledge of players.}) In more detail, we associate with $\DAO$ a function $\leakage : \algSpace \times \stateSpace \to \{0, 1\}^* \cup \{\bot\}$, which defines information about $\alg(\state)$ that can be predicted in advance of state $\state$. The following are a few examples of public functions:

\begin{newenum}
    \item \emph{Full privacy}: \emph{no information} about the investment strategy is revealed: 
    \[\leakage_\fullPrivFcnLabel(\alg, \state) = \bot.\]

    \item \emph{Asset transparency}: only the \emph{assets} involved in trades are revealed:
    \begin{align*}
            \leakage_\assetFcnLabel&(\alg, \state) = \big\{ \\
            &\asset_1, \asset_2\ \big|\ (\DAO, \state, \asset_1, \asset_2, N) \getsr \alg(\state)\big\}.
    \end{align*}    
    \iffull
    Note that $\leakage_\assetFcnLabel$ returns an \emph{unordered} set of assets, and thus does not reveal which is the asset being sold and which is the one being bought.
    \else
    \fi

    \item \emph{Directional transparency}: both the \emph{direction} of and \emph{assets} involved in trades are revealed:
    \begin{align*}
        \leakage_\directionFcnLabel&(\alg, \state) = \big( \\
        &\asset_1, \asset_2\ \big|\ (\DAO, \state, \asset_1, \asset_2, N) \getsr \alg(\state)\big).
    \end{align*}
    Unlike $\leakage_\assetFcnLabel$, $\leakage_\directionFcnLabel$ returns an \emph{ordered} tuple of assets, where the first element is the asset being purchased and the second is the asset being 
    \iffull
    sold, thus revealing the trade's direction.
    \fi
    
    \item \emph{Full transparency}: the \emph{entire} investment strategy is revealed:
    \[\leakage_\fullTransparencyFcnLabel(\alg, \state) = \alg(\state).\]
\end{newenum}

Public functions serve as theoretical abstractions to model the behavior of \sysnameshort revealed to other system players. For example, as mentioned earlier, $\leakage_\fullTransparencyFcnLabel$ could arise from white- or black-box access to
\iffull
$\alg$, or even more contrived scenarios like past trades from $\DAO$ leaking future trades that will be made.
\fi
Also notice that, depending on whether $\alg$ is deterministic or randomized, so is $\leakage$.

Equipped with public functions, we can now define the privacy of $\DAO$ as the ``distance'' between $\leakage(\alg, \cdot)$ and $\alg$ itself: a distance of 0 implies that $\DAO$ is fully transparent, while a maximum distance implies $\DAO$ is fully private. We proceed to make this definition concrete. 

First, note that $\leakage(\alg, \state)$ induces a probability distribution $\pi^\leakage_{\state}$ over $\tradeSpace$ for state $\state$, which represents the player's belief of the trade $\DAO$ may make at a given state. 
\iffull
For example, in the fully opaque case, $\pi^\leakage_{\state}$ is uniformly distributed over all possible trades $\DAO$ can make given the current market conditions and state of $\DAO$ (assets held, etc); in the asset transparency case, $\pi^\leakage_{\state}$ is uniformly distributed only over the trades that involve the two assets that are revealed; and so on. Based on this, we can define the privacy of $\DAO$ as the \emph{statistical distance} between distributions $\pi^\leakage_{\state}$ and $\pi_\state \gets \algDist(\state)$ (which, recall, is $\DAO$'s ``true'' distribution of trades). 
\fi
More formally:

\begin{definition}[Privacy of \sysnameshort]\label{def:privacy}
Let $\DAO$ be a \sysnameshortsingle with associated public function $\leakage$, $\pi^\leakage_{\state}$ be the probability distribution over $\tradeSpace$ induced by $\leakage$ for a state $\state$, and $\pi_\state \gets \algDist(\state)$ be the real distribution over $\tradeSpace$ for $\DAO$ for a state $\state$. We say that $\DAO$ is \emph{$\epsilon$-private} if, for any finite subset of states $\stateSpaceSubset \subseteq \stateSpace$,
it follows that
\[\frac{1}{|\stateSpaceSubset|}\sum_{\state \in \stateSpaceSubset} \delta(\pi^\leakage_{\state}, \pi_\state) \geq \epsilon\]
\iffull
where $\delta$ denotes the total variation distance between the pair of input random variables.
\fi
\end{definition}
In other words, the ``farther apart'' public information is from real information, the more private $\DAO$ is.
\iffull
We note that, while we use total variation distance, our definition of privacy is compatible with other standard statistical distance measures. 
\else
\fi

\section{Private CoinAlgs: Threats to Fairness}\label{sec:privacy}
Equipped with our formal model for \sysnameshort, we now begin our analysis of the \tradeoffName. 

As noted in Section~\ref{sec:what-are-coinalgs}, \sysnameshort in practice tend to be \emph{private}.
\iffull
This makes sense, since it is natural to keep investment strategies private in order to protect profits. 
\else
\fi
However, in this section, we will show that private \sysnameshort pose \emph{economic risks} to its investors. Intuitively, private \sysnameshort lead to \emph{information asymmetry}: privileged entities (e.g., a \sysnameshortsingle's creator) may have full access to the \sysnameshortsingle's investment strategy, which can be leveraged to \emph{extract value} from the \sysnameshortsingle's trades.

\begin{keyInsight}\label{keyinsight:opacity}
Investors in a \textit{private} \sysnameshortsingle $\DAO$---one whose investment strategy is not publicly disclosed---are vulnerable to unfair \emph{value extraction} by an adversary $\advA$ with insider knowledge of $\DAO$'s strategy.
\end{keyInsight}

\iffull
A core element of our analysis will be to frame economic (un)fairness in \sysnameshort as a form of value extraction, similar to MEV~\cite{babel2023clockwork,daian2020flash,qin2022quantifying,zhou2023sok} in other areas of decentralized finance, from which economic risks naturally follow. 
\else
\fi

\subsection{Economic Fairness Definition for \sysnameshort}
We begin by introducing our definition for economic (un)fairness of \sysnameshort. 
\iffull 
Alongside our privacy definition from Section~\ref{sec:definition}, this definition will let us show that privacy is tied to economic risk.
\else
\fi

As described in Section~\ref{sec:definition}, players have partial knowledge of the trades $\DAO$ will make, which we model via privacy function $\leakage$. To define economic fairness, we now assume that there is some priviledged, adversarial player $\advA$ with \emph{additional} knowledge of $\DAO$'s trades (formally, with a distribution over trades that is closer to $\pi$ than $\pi^\leakage$). This could correspond to, for example, the creators of $\DAO$. We assume that $\DAO$ is \emph{fully transparent} to $\advA$, so $\advA$ can predict the distribution of trades in advance; $\advA$'s only source of uncertainty may come from $\DAO$'s randomness (e.g., sampled inside a TEE).

\paragraph{Economic fairness definition.}
\iffull
Our goal is to define what it means for $\DAO$ to be ``fair'', i.e., that no player can extract outsized benefits from $\DAO$'s trades. To build intuition, we start with a straw definition, which shows important elements of a robust notion of economic fairness.
\fi

A first attempt could be to say that $\DAO$'s trades cannot be manipulated maliciously. So, for example, $\DAO$
\iffull 
cannot siphon funds to an adversarial asset, nor can 
\fi 
be steered to make trades that would benefit a particular player. While a legitimate threat, this definition only captures \emph{directly benefiting} from $\DAO$'s trades, and does not consider \emph{external capabilities and knowledge} of players, which may allow for \emph{indirect} benefit in the form of insider trading. For example, if $\advA$ knows that $\DAO$ will make a particular trade at a certain time, $\advA$ could sandwich the 
\iffull
trade to steal profits from $\DAO$.
\fi
Instead, we want our definition to capture the \emph{total value} that can be obtained (directly and indirectly) from $\DAO$'s trades. 
\iffull 
Even if $\DAO$ is still profitable, an adversary should not be able to leverage privileged information to make outsized gains.
\else
\fi
This leads to the core idea of our definition, which is to cast it as \emph{value extraction problem}.

Formally, we model economic fairness via security game $\FairnessGame$, defined in Figure~\ref{fig:fairness-game}. The game is parametrized by an adversary $\advA$ and an oracle $\oracle$, which is a \emph{prediction oracle} that models potential side information that $\advA$ may have. The game is also parametrized by a player $\player$, with access to $\DAO$'s public function $\leakage$. The game proceeds as follows. For every epoch between 1 and $d$ (a parameter to the game), starting from some initial state $\state_0$ (also a parameter to the game), $\DAO$ produces a trade, and $\advA$ (using $\oracle$) produces two sets of trades. These trades then execute together---$\DAO$'s trade between the two sets of trades from $\advA$---transitioning the blockchain to its next state. Similarly, $\player$ (using $\leakage$) also produces two sets of trades, which are interleaved with a trade from $\DAO$ to produce the next state. Note that, after $\state_0$, $\advA$ and $\player$ operate over separate transcripts. Lastly, the game outputs 1 if $\advA$'s balance increased by more than some threshold $\alpha$ over $\player$'s balance.

\begin{figure}[t!]
    \centering
	\fpage{0.47}
	{
	   $\procedurev{\FairnessGame^{\advA, \oracle}_{\DAO, \player, \state_0, d, \alpha}}$\\
        $\state_{\advA, 0} \gets \state_0$ \\
        $\state_{\player, 0} \gets \state_0$ \\
        $\cfor i \in [d]:$ \\
            $\ind \trade_{\DAO, \advA} \getsr \alg_\DAO({\state_{\advA, i}})$ \\
            $\ind \block'_\advA, \block''_\advA \getsr \advA^{\oracle}({\state_{\advA, i}})$ \\
            $\ind \block_\advA \gets \block'_\advA \concat (\trade_{\DAO, \advA}) \concat \block''_\advA$ \\ 
            \medskip
            $\ind \state_{\advA, i+1} \gets \action(\block_\advA, \state_{\advA, i})$ \\
            $\ind \trade_{\DAO, \player} \getsr \alg_\DAO({\state_{\player, i}})$ \\
            $\ind \block'_\player, \block''_\player \getsr \player^{\leakage_{\DAO}}({\state_{\player, i}})$ \\
            $\ind \block_\player \gets \block'_\player \concat (\trade_{\DAO, \player}) \concat \block''_\player$ \\
            $\ind \state_{\player, i+1} \gets \action(\block_\player, \state_{\player, i})$ \\
        $v_\advA \gets \balance(\advA, \state_{\advA, d}, \ETH) - \balance(\advA, \state_{\advA, 0}, \ETH)$ \\
        $v_\player \gets \balance(\player, \state_{\player, d}, \ETH) - \balance(\player, \state_{\player, 0}, \ETH)$ \\
        $\creturn |v_\advA - v_\player| > \alpha$
	}
	\caption{Game for the economic fairness definition of \sysnameshort.}
	\label{fig:fairness-game}
\end{figure}

We define the advantage of adversary $\advA$ over \sysnameshortsingle $\DAO$ (with associated public function $\leakage$) with respect to oracle $\oracle$, player $\player$, initial state $\state_0$, duration $d$ and threshold $\alpha$ as
\[\Advantage{\FairnessGameShort}{\DAO, \oracle, \player, \state_0, d, \alpha}(\advA) = \Big|\Prob{\FairnessGame^{\advA, \oracle}_{\DAO, \player, \state_0, d, \alpha} = 1}\Big|.\]

We now arrive at our definition of economic fairness:

\begin{definition}[Economic Fairness of \sysnameshort]\label{def:fairness}
    A \sysnameshortsingle $\DAO$ is \emph{$(\alpha, t)$-fair with respect to oracle $\oracle$} if for any adversary $\advA$, state $\state \in \stateSpace$ and integer $d \in \mathbb{N}$, there exists an efficient algorithm $\player$ such that
    \[\Advantage{\FairnessGameShort}{\DAO, \oracle, \player, \state_0, d, \alpha}(\advA) \leq t.\]
\end{definition}

In other words, our fairness definition characterizes the power of $\advA$'s predictive oracle $\oracle$ in terms of how it allows $\advA$ to extract additional profits. 

As mentioned earlier in the section, we are most interested in the case where $\oracle$ \emph{perfectly predicts} $\alg$, which is a special case of Definition~\ref{def:fairness}. We refer to the value that $\advA$ can extract with a perfect prediction oracle as the \emph{full knowledge extractable value} for a given starting state and set of parameters.

\iffull
Definition~\ref{def:fairness} models the (un)fairness of $\DAO$ in a parametrized way, since it is stated in terms of a particular threshold difference of value extraction and an adversarial success probability So, more accurately, it captures a \emph{snapshot} of $\DAO$'s overall fairness. Therefore, our definition is meant to be used in a flexible way, with multiple applications of our definition for various $(\alpha, t)$. Conceptually, this yields a \emph{fairness curve}, which is the broader characterization of $\DAO$'s fairness.
\else
\fi

\subsection{Fairness vs Privacy in \sysnameshort}
We now proceed to the main result of this section, arguing that privacy in \sysnameshort is in tension with economic fairness. In particular, we show that the existence of a meaningful information oracle (i.e., one which allows the adversary to extract additional value) implies that $\DAO$ must be private, and thus \emph{privacy is a precondition for unfairness}.

\begin{theorem}
    Let $\DAO$ be a \sysnameshortsingle that is $(\alpha, t)$-unfair with respect to perfect-prediction oracle $\oracle$ for some $\alpha, t > 0$. Then, there exists some $\epsilon > 0$ such that $\DAO$ is $\epsilon$-private.
    \label{thm:unfair_implies_private}
\end{theorem}
We defer the proof of this theorem to Appendix~\ref{sec:A-unfairness-privacy-proof}. Intuitively, if $\advA$ is able to extract additional value using $\oracle$ over any other $\player$ using $\leakage$, it must be the case that there is an information asymmetry between $\oracle$ and $\leakage$ that leads to this difference in value, which, by definition, implies that $\DAO$ is private. More formally, if $\advA^\oracle$ extracts additional value over $\player^\leakage$, a \emph{distinguisher} $\distinguisher$ attempting to differentiate $\oracle$ from $\leakage$ (which is an equivalent formulation of Definition~\ref{def:privacy}) can run both $\advA$ and $\player$ as subroutines, using the difference in extracted value to determine which oracle it is interacting with.

Note that Theorem~\ref{thm:unfair_implies_private} is a \emph{directional} result. An interesting question for future work is to study fine-grained variants, quantifying the degree to which privacy degrades for specific parameter values of Definition~\ref{def:fairness}. 

\medskip \noindent\textbf{Does privacy imply unfairness?}
A natural question to ask next is whether the converse of Theorem~\ref{thm:unfair_implies_private} also holds, i.e., whether privacy always leads to economic unfairness. This, however, need not always be true: additional information \emph{does not always lead to additional value extraction}, since information is not always useful from an economic perspective. 
The converse of Theorem~\ref{thm:unfair_implies_private} is therefore not true in general.

However, if it \emph{is} the case that $\DAO$ is private in such a way that privileged information is economically useful, then this directly leads to economic unfairness. This implication is by design, since our $\FairnessGame$ game is defined to precisely model the financial utility of information oracles. Therefore, an oracle from which more value can be extracted will (tautologically) lead to a larger value gap in our $\FairnessGame$ game.

Let us now formalize this idea. We model the \emph{financial utility} of an information oracle as the \emph{percentage of the full knowledge extractable value} (as defined earlier) that it enables. By definition, the perfect-prediction oracle $\oracle$ has a financial utility of 100\%. We then have the following simple result.

\begin{theorem}\label{thm:private_implies_unfair}
    Let $\DAO$ be a \sysnameshortsingle with associated public function $\leakage$ with a financial utility of $u$. Assume that $\DAO$ is $\epsilon$-private with respect to perfect-prediction oracle $\oracle$ for some $\epsilon >0$. Then, there exists some $\alpha, t>0$ such that $\DAO$ is $(\alpha, t)$-unfair with respect to perfect-prediction oracle $\oracle$.
\end{theorem}
As mentioned earlier, this result holds by construction: if $\leakage$ has a financial utility of $u$, $\player^\leakage$ will be able to extract at most $uT$ value in $\FairnessGame$, where $T$ is the full knowledge extractable value of $\alg$. On the other hand, $\advA^\oracle$ will be able to extract the value of $T$ in $\FairnessGame$.

\iffull
We note that the precise value of information is highly context-dependent. Therefore, deriving more general, fine-grained bounds on the degree to which (useful) information degrades economic unfairness is likely not feasible.
\else
\fi

\section{Transparent \sysnameshort: Threats to Profitability}
\label{sec:transparency}

Given our results in the previous section on the risks of economic unfairness in \textit{private} \sysnameshort, a better alternative would seem for \sysnameshort to be \textit{transparent}.
Transparency is, after all, a key property for ensuring the trustworthiness of systems ranging from open-source software to smart contracts. 

In this section we show, however, that full transparency in a \sysnameshortsingle---i.e., public knowledge of $\alg$---\textit{comes at the expense of profitability}. The reason is intuitive: Transparency means signaling trades publicly to the market. Arbitrageurs can profit from this advance information at the expense of the \sysnameshortsingle. Broadly, our work in this section provides evidence for: 
\begin{keyInsight}\label{keyinsight:transparency}
A \sysnameshortsingle $\DAO$ whose strategy $\alg$ is \textit{transparent}---known to an arbitraging adversary $\adv$---incurs a \textit{cost of transparency}: It yields lower profit than an equivalent \textit{private} \sysnameshortsingle $\DAO^*$. 
\end{keyInsight}

By equivalent \textit{private} \sysnameshortsingle $\DAO^*$, we mean that $\DAO^*$ executes the same strategy $\alg$ as $\DAO$, but with $\alg$ unknown to $\adv$ and thus not subject to frontrunning / sandwiching. $\DAO$'s \textit{cost of transparency} is the amount by which its profit is lower than that of $\DAO^*$.
\iffull
(Details follow below.)
\else
\fi

Even a small cost of transparency can place $\DAO$ at a competitive disadvantage. This disadvantage can lead to the market failure of $\DAO$, as investors gravitate toward the higher returns of a private alternative. As a result, Key Insight~\ref{keyinsight:transparency} strongly motivates the appeal of private \sysnameshort for profit maximization. Combined with our results in~\Cref{sec:privacy} regarding the risks of private \sysnameshort (Key Insight~\ref{keyinsight:opacity}), it is clear that \sysnameshort present significant risks however they are designed.

\subsection{Adversarial Strategies}

An adversary $\adv$ poses two distinct strategic threats to a transparent \sysnameshortsingle $\DAO$: 
\iffull

\bigskip
\noindent \textbf{Threat 1  (Strategy theft)}: If $\DAO$ is about to execute a trade $\trade_1$, $\adv$ can \textit{copy and frontrun} it, executing an equivalent trade $\trade_0$ prior to $\trade_1$. In this way, $\adv$ \textit{steals} $\DAO$'s investment strategy.

\bigskip
\noindent 
\textit{Example:}
Predicting a price rise for TOK from 10 now to 15 tomorrow, $\DAO$ prepares to buy 500 TOK in trade $\trade_1$. $\adv$ exploits $\DAO$'s transparency by  executing  $\trade_0 = \trade_1$ \textit{before} $\trade_1$---profiting if $\DAO$'s prediction was correct and potentially eroding the profit of $\adv$, as $\trade_0$ will raise the price of TOK paid by $\DAO$.  

\bigskip
\noindent \textbf{Threat 2  (Sandwiching)}:   $\adv$ can exploit $\DAO$'s transparency by  \textit{sandwiching} $\DAO$'s trade $\trade_1$---executing paired \emph{frontrunning} and \emph{backrunning} trades. In this way, $\adv$ profits from advance knowledge of price movement induced by $\tau_1$ while degrading $\DAO$'s profit.
    
\bigskip
\noindent
\textit{Example:} Given the same $\tau_1$ as above (buy 500 TOK), $\adv$ exploits $\DAO$'s transparency by executing $\trade_0$, buying 5 TOK before $\trade_1$, and selling 5 TOK in $\trade_2$ after $\trade_1$. $\adv$ profits because $\trade_1$ raises the price of TOK before $\adv$ sells, while $\trade_0$ raises the price of TOK that $\DAO$ pays in $\trade_1$, degrading $\DAO$'s profit.

\bigskip
\fi

Strategy theft is simple for an adversary to execute, as it just involves copying. However, it entails a hard-to-quantify risk for $\adv$, as $\DAO$'s prediction may be inaccurate and thus its strategy may or may not be profitable over any given time horizon. We perform experiments in~\Cref{sec:empirical_transparency} to evaluate the impact of strategy theft.

We focus in the remainder of this section on Threat 2. Sandwiching strategies are more amenable to analytic study than strategy theft. They can be  {\em risk-free}, in the sense that they do not depend on the soundness of $\DAO$'s strategy and instead yield immediate profit for $\adv$. Sandwiching is for this reason a widespread strategy today as a means of obtaining MEV~\cite{babel2023clockwork,daian2020flash,zhou2021high,zhou2023sok}. 

\iffull
\bigskip
\noindent
\textbf{Does sandwiching affect only transparent CoinAlgs?} Sandwiching---and broader MEV extraction---can degrade the profit of any transaction exposed to arbitrageurs, whether in public or private mempools. So even a private CoinAlg \textit{could} be vulnerable. A private CoinAlg \textit{can}, though, take advantage of sandwiching protections: services such as Flashbots Protect~\cite{flashbots_protect_rpc} or Cowswap~\cite{cowswap}, or side agreements with validators. Such protections are \textit{not} possible if $\DAO$ is transparent, as the trades executed will leak to $\adv$. 
\else
\fi

\subsection{Model Details}
\iffull
We first introduce a formal model for sandwich attacks against transparent \sysnameshort, extending our formalism from Section~\ref{sec:definition}. This model will form the basis for our results that follow.
\else
\fi

\paragraph{Market model.}
We consider a simple market in which two assets trade within a single, constant-product AMM. For the sake of concreteness, we denote these assets by TOK and USD (e.g., ether and a generic USD stablecoin). We let $t$ denote logical time for the AMM---a counter on the number of trades it executes.  The state of the AMM prior to execution in timestep $t$ (or after timestep $t-1$) is represented by a trading-pair balance $\balAMM_t = \pairbal{p_tr_t}{r_t}$.  In a constant-product AMM, this reflects a market TOK-USD price of $(p_tr_t) / r_t = p_t$. Our model of time matches that in~\Cref{sec:privacy}, but assumes that trades involve only the TOK-USD pair on this single AMM. In what follows, we use the term \textit{transaction} to denote the full, digitally signed payload that executes a \textit{trade} $\tau$ at the application layer, meaning an AMM operation. 

Trades are of two types: they either buy or sell TOK against USD. To simplify  notation from Section~\ref{sec:privacy}, we let $\buy{x}$ denote purchase of $x$ TOK and $\buy{-x}$ denote the sale of $x$ TOK. We assume AMM or smart-contract support for \textit{conditional trade execution} based on the TOK price in the AMM, i.e., support for limit orders. We denote a limit order with limit $\plim$ by $\buyl{x}{\plim}$, meaning that the trade will be executed as long as TOK's price is at or below $\plim$.\footnote{AMM trades generally specify not price limits, but \textit{slippage} limits. Price limits simplify our analysis and can easily be enforced via smart contract, as in our experiments in~\Cref{sec:empirical_transparency}.}

\paragraph{Model for $\DAO$'s strategy.} 
We are interested in algorithms $\alg$ for $\DAO$ with market-beating performance---motivating investor interest in the \sysnameshortsingle. We consider a simple model in which an internal \textit{oracle} $\oraclehat[\state_i]$ \textit{issues a prediction of the TOK-USD price} $\hat{p}$ for some future time $\hat{t}$.\footnote{$\oraclehat$ may reference off-chain data. As we assume that $\alg^\oraclehat$ is transparent, inputs to $\oraclehat[\state_i]$ are publicly available and canonically indexed by $i$ so that any player--- specifically $\adv$---can compute $\oraclehat[\state_i]$.} We let $\alg^{\oraclehat}$ denote use of this oracle in $\alg$. (For visual clarity, where appropriate we write $\alg^{\oraclehat}(\state_i)$ or even $\alg^{\oraclehat}$, rather than $\alg^{\oraclehat[\state_i]}(S_i)$.) It is $\oraclehat$, the source of the prediction $\hat{p}$, that embodies $\DAO$'s market-beating strategy---its ``secret sauce.''

We assume for simplicity here that $\alg^{\oraclehat}$ is \textit{deterministic} in $S_i$. We discuss possible relaxations of this assumption later. Furthermore, we assume, without loss of generality, that $\hat{p}$ is higher than the current TOK price in USD. (If $\hat{p}$ is lower, then $\DAO$ can treat USD as the target asset with the higher future price.)

$\DAO$'s strategy is a simple one-sided one. Its goal is to maximize the USD value of this portfolio by repeating the following procedure. Given current price $p$ and prediction $\hat{p} > p$ for TOK at future time $\hat{t}$:
\iffull
\begin{itemize}
    \item Execute $\buyl{u}{\plim}$ for TOK, where $u$ pushes the AMM price up to predicted price $\hat{p}$.
    \item Sell $u$ units of TOK at time $\hat{t}$.
\end{itemize}

Looking ahead, our empirical analysis in Section~\ref{sec:empirical_transparency} will consider even more sophisticated trading strategies.

For simplicity, we assume that sale of TOK perfectly realizes $\DAO$'s predicted market price, i.e., for TOK-USD trades, we disregard  slippage, frontrunning, etc.

\fi 

\paragraph{Transparency model.} As defined in~\Cref{sec:definition}, $\DAO$ is \textit{transparent} if every player, including $\advA$, has access to public function $\leakage_\fullTransparencyFcnLabel(\alg^\oraclehat, \cdot) = \alg^\oraclehat(\cdot)$. Recall that this definition models unlimited \emph{query access} to $\alg^\oraclehat(\cdot)$, but we make no assumptions about how this access realizes itself in
\iffull 
practice; $\leakage_\fullTransparencyFcnLabel$ is an abstraction for any setting that provides (at minimum) black-box access to $\alg^\oraclehat(\cdot)$. This could arise from access to some interface that $\advA$ can query, an open-source implementation of $\DAO$, etc.
\fi

A consequence---given our assumption that $\DAO$ is deterministic---is that  $\advA$ can \emph{predict trades from} $\alg^\oraclehat$ \emph{prior to their execution}. 
Exactly what ``prior to'' means will depend on the specific scenario considered in our game definitions and experiments. For example, looking ahead, our game $\sggame$, assumes that $\adv$ can learn $\tau_1$ in time to mount a successful sandwich attack against $\DAO$. In practice, for most blockchains, this means that $\adv$ can learn $\tau_1$ at the same time as $\DAO$ or even slightly later, but position its own transaction executing trade $\tau_0$ earlier in a block, as happens in MEV extraction today. This capability is a direct consequence of transparency, as $\adv$ can just execute $\alg^\oraclehat$ in parallel with $\DAO$. 

\paragraph{Model for adversary $\adv$.} We model $\adv$ monolithically, i.e., as a single entity. In practice $\adv$ could represent a collection of cooperating or independent arbitrageurs. We are interested in how $\DAO$ loses profit, and thus this is sufficient for our results; the way in which individuals profit is an orthogonal consideration. For a pair of sandwiching trades $(\tau_0, \tau_2)$ of $\adv$, our monolithic model gives an upper bound on $\adv$'s profits, as competition among arbitrageurs may degrade $\adv$'s profit (e.g., via MEV-auction cost). Our model yields a tight bound on $\DAO$'s losses due to sandwiching, however, as arbitrageurs will collectively maximize extracted value from $\DAO$ by pushing the AMM to $\DAO$'s limit price $\plim$ with $\tau_0$.\footnote{In an MEV supply chain, $\DAO$ might bid for advantageous block positioning to prevent sandwiching by $\adv$. $\DAO$ would need to bid up to $s - \epsilon$ against $\adv$, though, where $s$ is $\adv$'s sandwiching payoff. Our model excludes this possibility in the interest of a blockchain-agnostic analysis, but note that it would lead to the same high cost of transparency we show in our model.} We denote $\adv$'s portfolio prior to the execution of timestep $t$ by $\balAdv_t = \pairbal{c_t}{d_t}$.

\iffull
To gauge how a sandwich attack affects the profits of (transparent) $\DAO$, we compare it with a setting in which a \emph{private} \sysnameshortsingle $\DAO^*$ also executes the same strategy $\alg$; that is, $\advA$ instead has access to public function $\leakage_\fullPrivFcnLabel(\alg, \cdot) = \bot$, and so $\adv$ does not know the trades from $\DAO$ ahead of time. Therefore, we model this private setting by assuming no sandwiching by $\adv$. (Implicit here is also the assumption that $\DAO^*$ submits its  transactions so that they are not exposed in the mempool---as is possible in practice using services such as, e.g., Flashbots Protect~\cite{flashbots_protect_rpc}.)  More formally:

\begin{definition}[Cost of transparency]
    Let $\Pi_{\DAO}$ denote the payoff of transparent \sysnameshortsingle $\DAO$ in a given game at equilibrium, and $\Pi_{\DAO^*}$ denote that of its private equivalent $\DAO^*$. We define $\DAO$'s \textit{cost of transparency} as 
    $\Pi_{\DAO^*} - \Pi_{\DAO}$.
\end{definition}

That is, we use the setting with no sandwiching as ground truth, and compare $\DAO$'s payoffs in different settings in terms of how much they deviate from $\DAO^*$'s.
\else
\fi

\iffull
\paragraph{Blockchain model.} Block construction and infrastructure and markets around it vary considerably across blockchains and design evolutions. For generality, we abstract away block construction and just assume advantageous ordering of transactions by $\adv$---whether through use of MEV supply chains, exploiting $\DAO$'s transparency, etc.
\fi

\iffull
\subsection{Remarks on Modeling}

Before proceeding to our results, we make a few clarifying remarks about our model.

\begin{remark}[Trade vs.~transaction leakage]
The \textit{trade} in $\trade_1$, i.e., AMM operation, leaks to $\adv$ in our transparency model of~\Cref{sec:definition}. The \textit{transaction} $\trade_1$ itself, with its digital signature, may not. (In natural architectures, $\DAO$'s wallet would be independent of $\alg^\oraclehat$.) So in MEV supply chains, $\adv$ may not be able to execute the sequence $(\trade_0, \trade_1, \trade_2)$ atomically through bundling. \adv~\textit{can} still sandwich $\DAO$ non-atomically, which is roughly equivalent to atomic sandwiching in our monolithic model for $\adv$.
\end{remark}

\begin{remark}[(Weak) transparency assumptions]
\label{remark:prediction}

In practice, under $\leakage_\fullTransparencyFcnLabel$, the ability to predict trades prior to their execution is plausible under \textit{weak assumptions} on the power of the adversary: $\adv$ can learn $\tau_1$ in advance with only black-box access and \textit{need not reverse-engineer} $\alg^\oraclehat$, i.e., need not have white-box access. 

If $\DAO$ generates trades based on real-time information, $\adv$ can perform a sandwiching attack by executing $\alg^\oraclehat$ in parallel with $\DAO$ as noted above. Alternatively, if $\DAO$ makes medium-to-long-term price predictions, locks in a trade $\trade_1$ and then awaits favorable execution conditions, extraction of $\tau_1$ is possible by simulating plausible future blockchain states and emitting $\trade_1$ as the trade output consistently by $\alg^\oraclehat$ across simulations. 

Black-box access is of course possible with full disclosure of $\alg^\oraclehat$, but is possible even with just  (extensive) query access. 
\end{remark}

\begin{remark}[Determinism vs. non-determinism] While we assume here that $\DAO$ is deterministic---a strong assumption---we posit that our results also hold directionally for many natural embodiments of non-deterministic $\DAO$. The reason is that the set of maximally profitable trades for $\DAO$ is limited in scope. A profitably trading $\DAO$, for instance, would be unlikely to choose randomly between \textit{buying} TOK and \textit{selling} TOK. It would also be unlikely to choose randomly among a large set of assets or to vary trade sizes considerably.\footnote{Decomposing large trades into small ones is a common tactic for optimizing execution, but slippage, fees, etc., constrain the strategies.} So $\adv$ may not be able to predict $\tau$ \textit{exactly}, but can plausibly \textit{approximate} it well, which is sufficient to exploit it. \Cref{remark:prediction} applies also to non-deterministic $\DAO$ in this sense. We validate this claim empirically in Section~\ref{sec:empirical_transparency}, where we show in our experiments how $\advA$ is able to significantly degrade the profits of $\DAO$, even with imperfect knowledge of $\trade$.

Randomization of trade choices within profitable parameters may offer \textit{some} defense against exploitation by $\adv$---the basis for our proposed heuristic guardrail in~\Cref{sec:guardrails}.
\end{remark}
\fi

\subsection{Game-Theoretic Sandwiching Model}
\begin{figure}[t!]
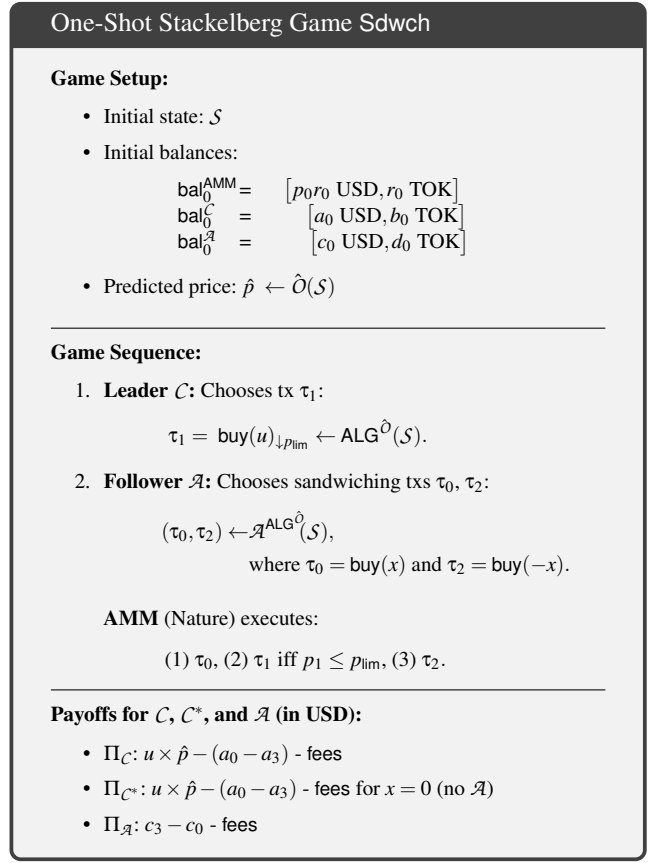

\begin{tcolorbox}[title=One-Shot Stackelberg Game $\sggame$]
{\footnotesize
\noindent \textbf{Game Setup:} 

\begin{itemize}
    \item Initial state: $\state$
   
    \item Initial balances:
    \begin{quote}
    \begin{tabular}{lll} 
       $\balAMM_0$&\hspace{-4mm}=\hspace{-6mm}&  $\pairbal{p_0r_0}{r_0}$\\
       $\balD_0$ &\hspace{-4mm}=\hspace{-6mm}&\hspace{3mm}$\pairbal{a_0}{b_0}$\\
       $\balAdv_0$&\hspace{-4mm}=\hspace{-6mm}&$\hspace{3.4mm}\pairbal{c_0}{d_0}$ 
    \end{tabular}
    \end{quote}
    
\item  Predicted price: $ \hat{p} ~\leftarrow \oraclehat(\state) $
\end{itemize} 

\medskip
\hrule
\medskip

\noindent \textbf{Game Sequence:}

    \begin{enumerate}
  \item \textbf{Leader $\DAO$:} Chooses tx $\trade_1$:
    \begin{align*}
        \hspace{-15mm} \trade_1 =\ \buyl{u}{\plim} \leftarrow \alg^{\oraclehat}(\state).
    \end{align*}
    
    \item \textbf{Follower $\adv$:} Chooses sandwiching txs $\trade_0$, $\trade_2$:
    \begin{align*}
       \hspace{3mm} (\trade_0, \trade_2) \leftarrow
    &\adv^{\alg^{\oraclehat}}\hspace{-1mm}(\state), \\ &\mbox{where } \trade_0 = \buy{x} \mbox{ and }\trade_2 = \buy{-x} .
    \end{align*}
\end{enumerate}

\smallskip
\hspace{7mm}\textbf{AMM} (Nature) executes: 
\begin{align*}
   \hspace{-3mm} &\mbox{(1) } \trade_0 \mbox{, (2) } \trade_1\mbox{ iff }p_1 \leq \plim \mbox{, } \mbox{(3) } \trade_2. 
\end{align*}

\hrule
\medskip

\noindent \textbf{Payoffs for $\DAO$, $\DAO^*$, and $\adv$ (in USD):} 

\begin{itemize}
    \item  $\Pi_{\DAO}$: $u \times \hat{p} - (a_0 - a_3)$ - \textsf{fees} 
        \iffull
        \item  $\Pi_{\DAO^*}$:   $u \times \hat{p} - (a_0 - a_3)$ - \textsf{fees}  for $x=0$ (no $\adv$)
        \else
        \fi

    \item $\Pi_{\adv}$: $c_3 - c_0 $ - \textsf{fees}  
\end{itemize} 

}
\end{tcolorbox}
\caption{Two-move Stackelberg game $\sggame$, modeling the interaction between transparent \sysnameshortsingle $\DAO$, acting as \textit{leader}, and arbitrageur $\adv$ acting as \textit{follower}.
$\DAO$ chooses transaction $\trade_1$, which $\adv$ sandwiches. $\DAO$'s payoff $\Pi_{\DAO}$ is its marginal portfolio value: predicted USD value of purchased TOK minus USD cost (for $\DAO^*$, without sandwiching by $\adv$). $\adv$'s payoff is pure USD profit.}
\label{fig:Stackelberg_game}
\end{figure}

To evaluate the interaction between $\DAO$ and $\advA$, we model our sandwiching setting as a basic \emph{Stackelberg game}~\cite{Stackelberg2011} between $\DAO$ and $\adv$, which we call $\sggame$, and detail in Figure~\ref{fig:Stackelberg_game}. A Stackelberg game models the interaction between a \emph{leader} and a \emph{follower}, who move 
\iffull 
sequentially: the leader moves first, and the follower reacts. 
\fi
Such games thus provide a natural abstraction for modeling sandwich-style attacks. A subtlety in our setting is that, while $\adv$'s transaction $\trade_0$ is executed first, $\DAO$ nonetheless acts as the leader, because $\adv$ conditions its sandwich transactions on $\DAO$'s transaction $\trade_1$.

As mentioned earlier, in order to bound exposure to exploitation by $\adv$, $\DAO$ can make $\trade_1$ a limit order, with some limit $\plim$.
\iffull
$\DAO$ then purchases TOK up to the future price $\hat{p}$.
\else
\fi
The value of $\plim$ will be important for our equilibrium analysis. Since $\DAO$ is fully transparent, however, $\advA$ is aware of the value of $\plim$ beforehand, and so $\advA$ can react to this defense from $\DAO$ accordingly.

\paragraph{Single stage game.} In one repetition of $\sggame$, there is a natural strategy for $\DAO$: set the limit on $\trade_1$ as $\plim \leftarrow p_0 + \epsilonMin$, where $p_0$ is the TOK price before execution of $\trade_0$ and $\epsilonMin$ is the minimum price increment. 
\iffull
That is, $\DAO$ adopts a conservative defense strategy, aborting its trade as soon as an arbitrageur attempts to sandwich. 
\else
\fi
This strategy yields a Stackelberg equilibrium with optimal payoff for $\DAO$. (Setting $\plim = 0$ would enable $\adv$ to deny a payoff to $\DAO$ at zero cost, as in the repeated-game setting discussed later.) $\DAO$ captures nearly all surplus, because a sandwich attack by $\adv$ is viable only up to price $\epsilonMin$. We capture this in the following simple theorem:

\iffull
\begin{theorem}[$\DAO$'s cost of transparency in single-shot Game $\sggame$ is zero.]
    In the single-shot variant of $\sggame$, the unique equilibrium is when $\DAO$'s cost of transparency is zero, i.e., $\Pi_{\DAO^*} - \Pi_{\DAO} = 0$.
\end{theorem}
\fi
This simply follows from the fact that, since $\advA$'s transaction will increase the price by at least $\epsilonMin$, it cannot increase its payoff by sandwiching the trade.

\paragraph{Repeated game.} In practice, however, \sysnameshort and front-runners engage in long-term, multi-block interactions, which drastically change the equilibrium analysis. We thus now turn to the \emph{repeated} variant of $\sggame$. The key difference in this case is that, as is standard for repeated games, players can now use \emph{trigger strategies}~\cite{mailath2006repeated} to \emph{force cooperation}. Such strategies create a credible threat of \emph{punishment} in future rounds, thereby incentivizing cooperation in present rounds.

Our core insight in this case is that $\adv$ can \textit{weaponize transaction invalidation} as a \emph{grim-trigger} strategy. That is, $\adv$ can execute a transaction with trade $\tau_0$ that buys $x$ TOK and raises the USD price of TOK from current price $p$ to \textit{above} $\plim$ (or the inverse)---causing $\DAO$'s transaction executing $\tau_1$ to fail. Therefore, $\advA$ can threaten to invalidate \emph{all} future transaction from $\DAO$ and deny future profits if $\DAO$ does not raise $\plim$ and share the surplus, thereby forcing cooperation via the threat of punishment. (We note that we we focus our analysis on grim-trigger strategies for simplicity, but our results generalize to more fine-grained trigger strategies.)

The equilibrium under a trigger strategy will depend critically on the cost of transaction invalidation for the adversary. At most, $\adv$ loses the marginal value  of its TOK (equal to $x(\plim - p)$) to arbitrage / MEV-auction costs and pays blockchain transaction fees (essentially a constant $c$) plus AMM trading fees (a fixed percentage $\AMMfee$ of trade size, thus $\AMMfee\plim$). At a minimum, $\adv$ incurs the fees. We quantify this formally bellow:

\begin{lemma}
\label{lem:cost-buffer}
In $\sggame$, 
if the adversary~$\adv$ frontruns with
$\trade_0=\mathsf{buy}(x)$ far enough to raise the AMM price from
$p$ to $>p+\plim$ (so $\trade_1$ fails), $\adv$'s transaction cost is greater than
\begin{equation}
C(\plim)\;\coloneqq\;
p\, r_0\!\left(\sqrt{\frac{p+\plim}{p}}\;-\;1\right) \mbox{ USD}.
\label{eqn:C_beta}
\end{equation}

\end{lemma}
We show the proof of this lemma in Appendix~\ref{subsec:A-cost-buffer-proof}. Equation~\ref{eqn:C_beta} may be viewed as a \textit{penalty} imposed upon $\adv$ for invalidating a trade from $\DAO$.
\iffull
Therefore, there is a tradeoff between $\plim$ and payoffs in $\sggame$: the lower $\plim$ is, the more profit $\DAO$ may extract, but the cheaper it is for $\advA$ to invalidate the trade.
\else
\fi

Our goal is to analyze $\sggame$ under a grim-trigger strategy and show that $\advA$ can force an equilibrium where $\DAO$ must increase $\plim$. However, analyzing $\sggame$ directly is complex, since many exogenous factors (outside the scope of $\DAO$ and $\advA$'s interaction) are involved in state changes; a full analysis of $\sggame$ needs to take into account changes from this environment. Therefore, our approach will be to instead analyze equilibrium in a \emph{stylized variant} of $\sggame$, which abstracts away external factors and reduces the game to its key components.

\subsection{Stylized Sandwiching Model}
At its core, the interaction between $\DAO$ and $\advA$ involves an available surplus (due to $\DAO$'s price prediction), which $\DAO$ splits between itself and $\advA$ (by setting $\plim$). Then, $\advA$ can choose to accept or reject this split (by invalidating the transaction or not). This setting can naturally be modelled as an \emph{ultimatum game}~\cite{harsanyi1961rationality}. Such games, standard in the game theory literature, consider a \emph{proposer}, who suggests a division of a sum of money; and a \emph{responder}, who accepts or rejects the split. 
\iffull
This type of game maps directly to the abstract interaction between $\DAO$ and $\advA$.
\else
\fi
We show our stylized ultimatum game, $\ultgame$, in Figure~\ref{fig:ultimatum_game}. $\ultgame$ is simple: there is an available surplus $s$, which $\DAO$ splits at some cutoff point $e$, yielding $e$ to $\advA$ and keeping the remaining $s-e$ to itself. Then, $\advA$ can choose to accept the split or burn the entire surplus, indicated by its choosing of a bit $b$.

\begin{figure}[t!]
\begin{tcolorbox}[title=Ultimatum game $\ultgame$]
{\footnotesize

\noindent \textbf{Game Sequence:}

    \begin{enumerate}
  \item \textbf{Leader $\DAO$:} Divides surplus $s$ between itself and $\advA$:
    \begin{align*}
        &(e, s - e) \gets \DAO, \text{where } 0 \leq e \leq s
    \end{align*}
    
    \item \textbf{Follower $\adv$:} Accepts or denies split of surplus:
    \begin{align*}
       &b \gets \advA(e), \textnormal{where } b \in \{0,1\}
    \end{align*}
\end{enumerate}

\hrule
\medskip

\noindent \textbf{Payoffs for $\DAO$, $\DAO^*$, and $\adv$ (in USD):} 

\begin{itemize}
    \item  $\Pi_{\DAO}$: $bc_1(s-e) - \textsf{fees}$ 
    \item $\Pi_{\adv}$: $bc_2e - (1-b)z\sqrt{e} - \textsf{fees}$  
\end{itemize} 
for constants $c_1 > c_2 > 0$ and $z>0$.
}
\end{tcolorbox}
\caption{Ultimatum game $\ultgame$, which is a stylized variant of game $\sggame$. \sysnameshortsingle $\DAO$ offers a split of the available surplus to $\advA$, who can accept the split or deny profit to both (at a cost proportional to the piece it is offered).}
\label{fig:ultimatum_game}
\end{figure}

\paragraph{Payoff structure.} A challenge in defining our stylized game is determining the payoff structure, i.e., how the payoffs in the general $\sggame$ game translate to this simpler abstraction. In particular, notice that our setting differs from standard ultimatum games in that $\advA$ must 
\iffull
\emph{pay to reject the split}, since there is a cost associated with invalidating $\DAO$'s transactions (Lemma~\ref{lem:cost-buffer}); rejection generally leads to a payoff of 0 in ultimatum games, but not in our setting. 
\fi
Let us now analyze the payoffs when $\advA$ accepts the split and when it does not.

If $b=1$, $\DAO$'s trade is not canceled, and thus it can benefit from its future price prediction, while $\advA$ can extract the immediate value of the sandwich. This maps directly to our stylized model as a payoff of $(s-e)\hat{p}$ for $\DAO$ and $ep$ for $\advA$, which we can simplify as $c_1(s-e)$ and $c_2e$ for constants $c_1 \geq c_2 > 0$.

If $b=0$, $\DAO$ receives a payoff of 0, while $\advA$ pays the invalidation cost from Equation~\ref{eqn:C_beta}. Mapping Equation~\ref{eqn:C_beta} to our stylized game, $\plim$ corresponds to $p + e$, and so $\advA$'s cost to reject the split is
\[C(e) \coloneqq p\,r_0\!\left(\sqrt{\frac{2p+e}{p}}-1\right).\]
Notice that $C(e)$ scales with $\sqrt{e}$, and so we can simplify this expression as a function of $\sqrt{e}$:
\begin{lemma}\label{lem:cost-buffer-stylized}
    For any $e \geq 0$, there exists a constant $z$ such that
    \[C(e) \leq z \sqrt{e}\]
\end{lemma}

We show the proof of this lemma in Appendix~\ref{subsec:A-cost-buffer-stylized}. We use this upper bound on the invalidation cost since it allows us to simplify the equilibrium analysis, while considering a setting that is in fact more beneficial for $\DAO$.

\iffull
Putting it all together, the payoffs for $\DAO$ and $\advA$ in $\ultgame$ are:
\begin{align*}
    \Pi_{\DAO} \coloneqq bc_1(s-e) + (1+b)0 - \textsf{fees},\text{ and} \\
    \Pi_{\adv} \coloneqq bc_2e + (1-b)(-z\sqrt{e}) - \textsf{fees}.
\end{align*}
\fi
Notice how $b$ acts as a switch, controlling the payoff for each party depending on whether $\advA$ accepts the surplus or not.

\paragraph{Repeated game equilibrium.} We now analyze the repeated game variant of $\ultgame$, showing that $\advA$ can use a grim-trigger strategy to enforce an equilibrium where $\DAO$ shares the surplus. Critically, however, $\advA$'s grim trigger must be a \emph{credible threat} in order to force cooperation: if it is not rational for $\advA$ to employ the grim trigger, $\DAO$ will just keep the entire surplus and ignore the threat. However, a subtlety of our setting is that, as explained earlier, there is a cost associated with the grim-trigger, and so $\advA$ would in fact obtain a higher payoff by accepting \emph{any} split of $s$, however small. Therefore, a priori, employing a grim-trigger strategy would not be rational for $\advA$, and does not pose a credible threat. 

To address this, $\advA$ can use the standard technique of \emph{committing to using the grim-trigger before the game begins}. By using a \emph{commitment device}~\cite{kalai2010commitment} to bind usage of the grim-trigger, $\advA$ can influence $\DAO$ to behave as if the grim-trigger is a credible threat, even if deciding to use it \emph{after} $\DAO$'s defection is irrational. There is a rich body of work exploring ways to commit to strategies in various settings~\cite{bagwell1995commitment}; our results in this section are agnostic to the specific mechanism used by $\advA$. In particular, blockchains provide natural commitment devices via smart contracts~\cite{hall2021game}.

We now arrive at our main result, which is that $\advA$ can use a grim-trigger strategy to force $\DAO$ to share the surplus.

\begin{theorem}\label{thm:ult-game-eq}
    In the repeated variant of $\ultgame$, there exists an $e^* > 0$ such that $\advA$ can use a grim-trigger strategy to enforce an equilibrium where, in every round, $\DAO$ divides the surplus $s$ as $(e^*, s-e^*)$.
\end{theorem}

\iffull
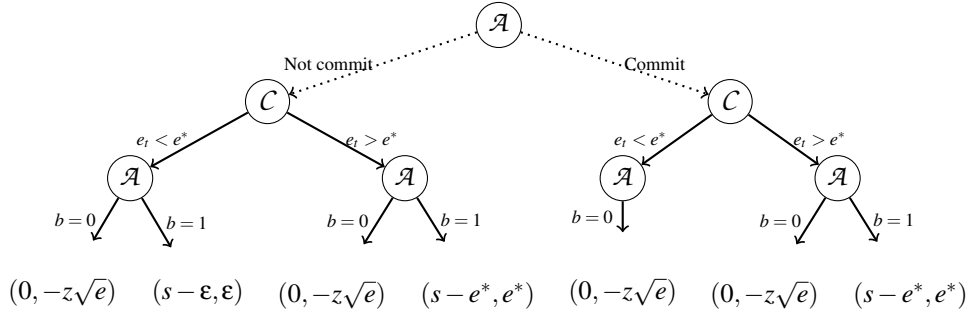
\begin{figure*}[t!]
    \centering
    \begin{forest}
    for tree={
        circle, draw,
        minimum size=0.4cm,
        inner sep=1pt,
        align=center,
        s sep=0.3cm,
        l=0.3cm,
        edge={->, thick},
        grow=south
    }
    [$\advA$
        [$\DAO$, edge={->, thick, dotted}, edge label={node[midway, left, font=\scriptsize]{Not commit}}
            [$\advA$, edge label={node[midway, left, font=\scriptsize]{$e_t < e^*$}}
                [{$(0, -z\sqrt{e})$}, edge label={node[midway, left, font=\scriptsize]{$b=0$}}, draw=none, circle=false]
                [{$(s-\epsilon, \epsilon)$}, edge label={node[midway, right, font=\scriptsize]{$b=1$}}, draw=none, circle=false]
            ]
            [$\advA$, edge label={node[midway, right, font=\scriptsize]{$e_t > e^*$}}
                [{$(0, -z\sqrt{e})$}, edge label={node[midway, left, font=\scriptsize]{$b=0$}}, draw=none, circle=false]
                [{$(s-e^*, e^*)$}, edge label={node[midway, right, font=\scriptsize]{$b=1$}}, draw=none, circle=false]
            ]
        ]
        [$\DAO$, edge={->, thick, dotted}, edge label={node[midway, right, font=\scriptsize]{Commit}}
            [$\advA$, edge label={node[midway, left, font=\scriptsize]{$e_t < e^*$}}
                [{$(0, -z\sqrt{e})$}, edge label={node[midway, left, font=\scriptsize]{$b=0$}}, draw=none, circle=false]
            ]
            [$\advA$, edge label={node[midway, right, font=\scriptsize]{$e_t > e^*$}}
                [{$(0, -z\sqrt{e})$}, edge label={node[midway, left, font=\scriptsize]{$b=0$}}, draw=none, circle=false]
                [{$(s-e^*, e^*)$}, edge label={node[midway, right, font=\scriptsize]{$b=1$}}, draw=none, circle=false]
            ]
        ]
    ]
    \end{forest}
    \caption{Game tree for strategy profile that leads to a beneficial equilibrium to $\advA$ in game $\ultgame$ (Figure~\ref{fig:ultimatum_game}). Equilibrium is at the rightmost node.}
    \label{fig:ult-game-eq}
\end{figure*}

\else
\fi

\iffull
We show the proof of this theorem in Appendix~\ref{subsec:A-transparency-proof}, including the game tree for the strategy profile in Figure~\ref{fig:ult-game-eq}.
\fi
Our proof is straightforward, using a standard argument for the analysis of repeated games: we evaluate the payoff in the cooperative path and when $\DAO$ deviates, showing that $\DAO$ achieves a higher payoff in the former due to $\advA$'s grim-trigger threat.

\section{Empirical Evaluation}
\label{sec:empirical_transparency}

In this section, we conduct experiments demonstrating the impact of frontrunning on backdoored \sysnameshort in a realistic setting using historical blockchain data. Through these experiments, we learn the extent to which frontrunning a \sysnameshortsingle is profitable even with limited, imprecise information. All experiment source code can be found at \url{https://git.disroot.org/verge24/calgs}.

\paragraph{Motivation.} We demonstrated the theoretical risk of profit capture in \sysnameshort in Section~\ref{sec:transparency}. We now aim to discover the potential relative magnitude of such capture under real market conditions. Are \sysnameshort actually at risk? Can a frontrunner profit with limited information?

\paragraph{Methodology overview.} Following our model from Section~\ref{sec:transparency}, we design a synthetic \sysnameshortsingle with perfect knowledge of future market conditions and simulate its trades on historical block data. It represents the ideal \sysnameshortsingle: its transactions are submitted privately, and investors cannot lose money as long as it is not blocked from trading. A simple frontrunner attempts to take advantage of its trades. These simulations reflect the liquidity conditions and fees present in an actual token market while showcasing the risk of frontrunning even for a best-case \sysnameshortsingle.

The experiments in this section complement the preceding theoretical results, to provide evidence for the following:

\begin{keyInsight}
    A \sysnameshortsingle that is profitable as a result of accurate market predictions can suffer significant losses in profit in the presence of even a low-bandwidth channel that leaks trade information to an arbitraging adversary.
\end{keyInsight}

As we unpack later in the section, using a special testing harness, we perform a variety of experiments to test this hypothesis across configurations of variables reflective of real-world considerations: multiple types of assets (stable and volatile), \sysnameshortsingle designs (transparent and partially private), adversarial strategies (strategy theft and sandwiching), and more. Broadly, our experiments show that, regardless of starting conditions, a \sysnameshortsingle that reveals even a small amount of information is at risk of profit loss.

\begin{figure*}[t!]
	\centering
	\begin{tabular}{%
			p{0.17\linewidth}   %
			p{0.09\linewidth}   %
			p{0.075\linewidth}  %
			p{0.11\linewidth}   %
			p{0.22\linewidth}   %
			p{0.087\linewidth}   %
			p{0.101\linewidth}   %
		}
		\toprule
		\textbf{\sysnameshortsingle type} &
		\textbf{Adversarial strategy} &
		\textbf{Asset} &
		\textbf{\sysnameshortsingle start balance} &
		\textbf{Adversary start balance} &
		\textbf{Start time offset} &
		\textbf{Execution window} \\
		\midrule
		\makecell[l]{%
			\begin{minipage}[t] {\linewidth}
				\raggedright
				\begin{itemize}[leftmargin=*]
					\item Fully transparent
					\item Transparent, secret trading times
					\item Private with one‑bit channel
				\end{itemize}
			\end{minipage}
		} &
		\makecell[l]{%
			\begin{minipage}[t] {\linewidth}
				\raggedright
				\begin{itemize}[leftmargin=*]
					\item Strategy theft
					\item Sand\-wiching
				\end{itemize}
			\end{minipage}
		} &
		\makecell[l]{%
			\begin{minipage}[t] {\linewidth}
				\begin{itemize}[leftmargin=*]
					\item USDC
					\item PEPE
				\end{itemize}
			\end{minipage}
		} &
		\makecell[l]{%
			\begin{minipage}[t] {\linewidth}
				\begin{itemize}[leftmargin=*]
					\item 12 ETH
					\item 60 ETH
					\item 300 ETH
					\item 1,500 ETH
				\end{itemize}
			\end{minipage}
		} &
		\makecell[l]{%
			\begin{minipage}[t] {\linewidth}
				\raggedright
				\small
				\begin{itemize}[leftmargin=*]
					\item 8 ETH / \mbox{16,000 USDC}
					\item 40 ETH / \mbox{80,000 USDC}
					\item 200 ETH / \mbox{400,000 USDC}
					\item 1,000 ETH / \mbox{2,000,000 USDC}
				\end{itemize}
			\end{minipage}
		} &
	\makecell[l]{%
		\begin{minipage}[t] {\linewidth}
			\begin{itemize}[leftmargin=*]
				\item 0 hours
				\item 6 hours
				\item 12 hours
				\item 18 hours
			\end{itemize}
		\end{minipage}
	} &
		\makecell[l]{%
			\begin{minipage}[t] {\linewidth}
				\begin{itemize}[leftmargin=*]
					\item 0.75M \small{blks}
					\item 1.5M blks
					\item 2.25M blks
					\item 3M blks
				\end{itemize}
			\end{minipage}
		}
	    \\
	    \bottomrule
	\end{tabular}
	\caption{Summary of the experimental parameters and the values we consider.  The testing harness evaluates the cross product of all values.}
	\label{fig:experiments-parameters}
\end{figure*}

\subsection{Experimental Setup and Methodology}

\paragraph{\sysnameshortsingle strategy.}

The \sysnameshortsingle trades on a predetermined schedule and will trade up to the future market price predicted at the time of its next scheduled trade to ensure a profit, using as many funds as it can. Our simulated \sysnameshortsingle uses a more sophisticated strategy than that reflected in our game-theoretic model of Section~\ref{sec:transparency}, which is simplified for analytic tractability. This \sysnameshortsingle uses continually updated information: it adjusts its strategy at every trading period using the next predicted price\iffull{ and does not necessarily sell the assets it purchased in the last trade. For example, if the \sysnameshortsingle predicts the price of ETH will increase by \$50 today, it will try to buy ETH now; if tomorrow the \sysnameshortsingle predicts another increase of \$70, it will try to buy more ETH tomorrow and will not sell its current ETH holdings. }\fi It submits its trades through a private transaction relay to protect against ordinary sandwiching attacks (i.e., the frontrunner does not observe its transactions before they are included in blocks).

A core goal of our experiments is to understand the extent to which a \sysnameshortsingle's profit loss depends on the \emph{amount} of information that is leaked: Is even a small amount of leakage sufficient for value extraction? We thus experiment with several variants that represent different types of leakage: (1) a fully transparent trading strategy that trades on a fixed interval, (2) a private strategy trading on a fixed schedule that leaks only a binary signal to the frontrunner (``buy/sell''), and (3) a transparent strategy with secret trading times but a known trading time distribution; in particular, for the third type, the \sysnameshortsingle places its trades according to a Poisson process, aiming to trade twice per day on average ($\lambda = 3{,}600 \text{ blocks}$). The aim of these three settings is to test how several plausible \emph{degrees} of information leakage affect a \sysnameshortsingle's profitability.

Using these three \sysnameshortsingle types as a starting point, we further vary lower-level details of the starting conditions to test whether these also bear an impact on our results. In particular, we experiment with different asset types, starting balances, times of day trades are executed, and the overall window of time in which the \sysnameshortsingle runs. See~\Cref{fig:experiments-parameters} for a summary of the parameters we vary and the values we consider. Our goal is to cast a wide net over \sysnameshortsingle types: those that trade volatile vs. stable assets, have large vs. small balances, execute for short vs. long periods of time, etc. As we will see, varying these parameters does not affect the directionality of our results.

\paragraph{Adversarial strategies.} Turning to adversaries, we similarly consider multiple strategies, as motivated by the threats from earlier sections: (1) strategy theft, where the frontrunner copies the \sysnameshortsingle's trading strategy; and (2) sandwiching, where the adversary, doubting the profitability of the \sysnameshortsingle's strategy, frontruns and backruns the \sysnameshortsingle's trades. 

For sandwich attacks, an important consideration is the amount of capital that the adversary should use in its transactions. While one might expect the best strategy is to use as much available capital as possible, this is not optimal, for two reasons: (1) larger trades incur higher pool fees, and (2) larger trades risk increasing asset prices above the \sysnameshortsingle's price limit. Therefore, given a particular \sysnameshortsingle trade to frontrun, an attacker's estimated sandwich profit is instead a \emph{unimodal function} of its trade input, with a peak that is determined by the two aforementioned conditions. We show an example of this in Figure~\ref{fig:fr-profit-single}, where we quantify the relationship between an attacker's profit and its sandwich input size.

Because this function is strictly unimodal, we can use a \emph{golden-section search} over frontrun trade sizes to optimize the frontrunner's trade input~\cite{kiefer1953minimax}. We use this technique to derive the frontrunner's sandwich transactions from its available capital.

Similar to the \sysnameshortsingle, we consider multiple values for the adversary's starting balance, in order to analyze the feasibility of profit extraction for both well-resourced and under-resourced adversaries. See~\Cref{fig:experiments-parameters} for a summary of the values we consider. As we discuss later, our simulations show that adversaries of both types are able to profit.

\iffull
\begin{figure}
    \centering
    \includegraphics[width=0.85\linewidth]{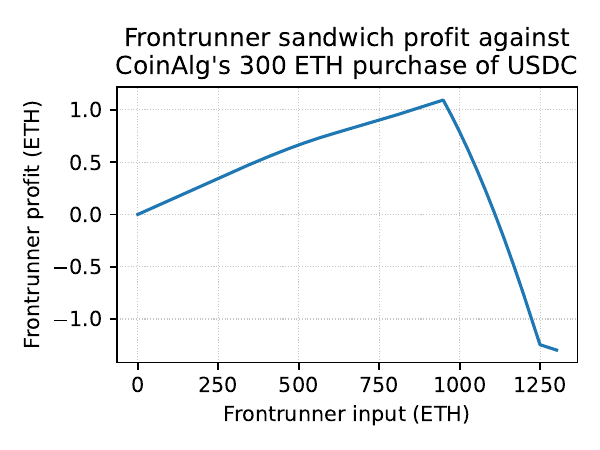}
    \caption{Profit from sandwiching a 300-ETH \sysnameshortsingle purchase of USDC in the 0.05\% Uniswap V3 pool at Ethereum block 20,000,000 varied by frontrunner trade size. Profit decreases once the \sysnameshortsingle begins to trade with less of its capital, adhering to its price limit. With a frontrunner input of 1,250 ETH, the \sysnameshortsingle does not place a trade.}
    \label{fig:fr-profit-single}
\end{figure}
\fi

\paragraph{Simulation details.}
We conduct experiments on two real-world Uniswap V3 token pools (USDC/ETH, 0.05\% fee and PEPE/ETH, 0.3\% fee) and target a trading period between Ethereum blocks 20,000,000 and 23,000,000 (June 1, 2024 to July 26, 2025). Using an archive node, we simulate trades by running a multi-transaction call at the top of a target block whenever a player wants to submit a transaction. The multi-call (1) funds the \sysnameshortsingle's and frontrunner's simulated accounts by transferring assets from burn addresses (ETH, PEPE) or using an authorized token minter (USDC), (2) approves and trades tokens on the target market, and (3) returns the transaction results and reports the final balances of the players.

To prevent these transactions from disrupting historical trading activity in the simulation, historical state is restored after each simulated block. We incorporate a third ``ghost'' player who places a trade at the beginning of every multi-call to compensate for the market movements caused by the \sysnameshortsingle and frontrunner in previous blocks.
We eliminate these market corrections after the \sysnameshortsingle has traded and the frontrunner has had the option to react to that trade.
This ensures the \sysnameshortsingle's price predictions remain accurate and token prices interpretable. Note that this negates the impact of a frontrunner's back-run: the back-run does not negatively affect the \sysnameshortsingle. Our experimental results thus show an upper bound on \sysnameshortsingle profits; frontrunning impact may be greater.
\iffull
Our simulations also differ from our model in that the \sysnameshortsingle is given a finite amount of funds.
\fi

Notice that, for a fixed set of starting conditions (i.e., selecting one value per column in Figure~\ref{fig:experiments-parameters}), the results of a simulation run are \emph{deterministic}. Therefore, a single run per parameter set is sufficient to obtain our results. The only exception is our experiments in which the \sysnameshortsingle decides when to trade according to a Poisson distribution. In this case, we performed 16 experimental runs with different random seeds for every combination of parameters.

\subsection{Results}
We ran our testing harness across all 3,072 combinations of parameters shown in Figure~\ref{fig:experiments-parameters}. The output of all experimental runs can be found at \url{https://git.disroot.org/verge24/calgs/src/branch/main/results}.

The primary takeaway from our experiments is that, across all parameter combinations, \emph{the presence of an adversary leads to profit loss for the \sysnameshortsingle}, confirming our hypothesis that a \sysnameshortsingle that reveals information (to either the public or to an insider through a backdoor) incurs a cost. While the specific amount of profit loss depends on the starting conditions (particularly the initial balances), the behavior is directionally consistent across all instances.

We now discuss a few lower-level takeaways elucidated by our experiments.

\paragraph{Both adversarial strategies are viable.} Our experiments confirm that, as expected, both strategy theft and sandwiching are profitable strategies for the adversary. 

\begin{figure}[t!]
    \centering
    \includegraphics[width=\linewidth]{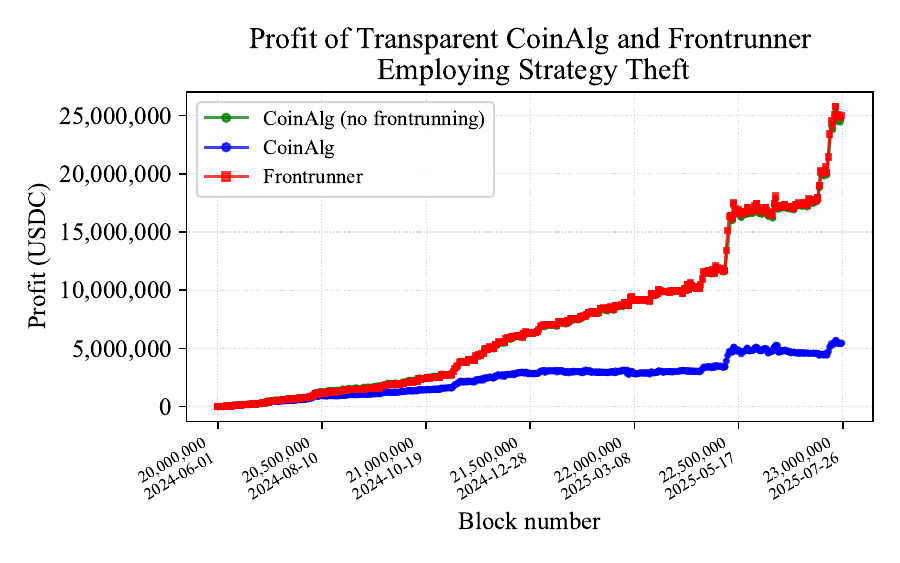}
    \caption{Experimental run showing the profit of the ideal CoinAlg (blue) starting with 300 WETH using a transparent trading algorithm in the WETH/USDC 0.05\% pool, frontrunner (red) starting with 200 WETH / 400,000 USDC employing strategy theft, and the same \sysnameshortsingle (green) if there were no frontrunner. Note that the frontrunner's profit and the \sysnameshortsingle's profit without frontrunning in the graph are nearly identical.}
    \label{fig:fr-profit-strategy-theft}
\end{figure}

For strategy theft, the frontrunner copies each \sysnameshortsingle trade and submits a transaction one block before the \sysnameshortsingle does, getting a better price and reducing the remaining liquidity available within the desired price limit. We exemplify this in~\Cref{fig:fr-profit-strategy-theft}, which shows the result of one of our experimental runs that consider a transparent \sysnameshortsingle and a strategy-theft adversary. As we can see, the frontrunner captures nearly all the profit the \sysnameshortsingle would have captured in the absence of frontrunning.
With sufficient capital, its larger market trades consume available liquidity and reach the target price limit, leading to more variable profit for the frontrunner and significantly hampered trading ability for the \sysnameshortsingle. The \sysnameshortsingle is blocked from trading when the frontrunner is able to move the price to the price limit.

\begin{figure}[t!]
    \centering
    \includegraphics[width=\linewidth]{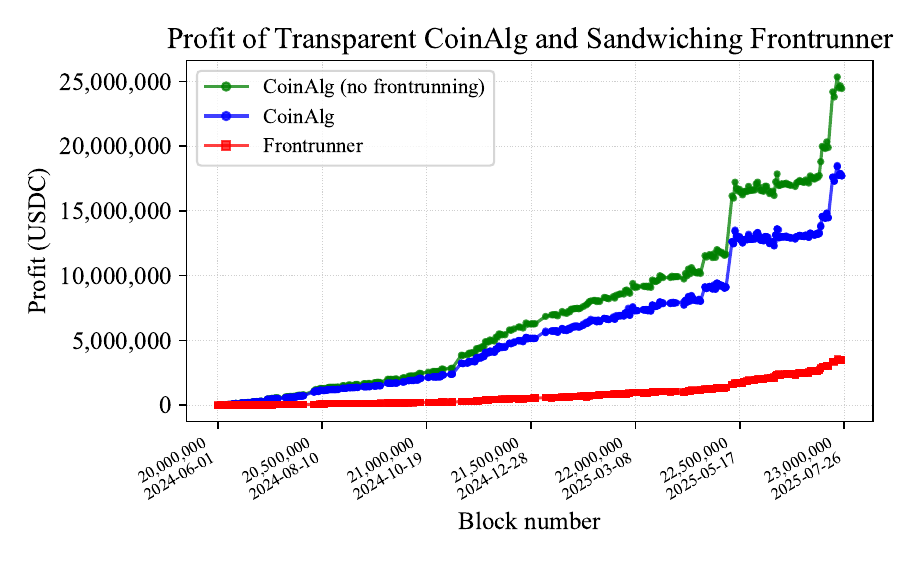}
    \caption{Experimental run showing the profit of a transparent \sysnameshortsingle (blue), a sandwiching frontrunner (red), and the \sysnameshortsingle if there were no frontrunner (green), using the same experimental configuration as~\Cref{fig:fr-profit-strategy-theft}.}
    \label{fig:fr-profit-transparent-sandwiching}
\end{figure}

Traditional sandwiching activity usually involves positioning trades around a victim trade in the same block. \sysnameshort using anti-frontrunning best practices, such as private mempools, can still be frontrun with sandwich attacks if their trades can be predicted. As exemplified in~\Cref{fig:fr-profit-transparent-sandwiching}, frontrunner sandwich attacks reduce the profits of our transparent \sysnameshortsingle by 27.6\%. Our ideal \sysnameshortsingle is still profitable despite sandwich activity because the price limits it uses guarantee that trades are profitable and are not based on ``slippage'' relative to the market price; furthermore, frontrunners cannot impact the \sysnameshortsingle when its action is to hold assets rather than trade them.

\paragraph{Even partially-private \sysnameshort are vulnerable.} Our experiments also show that full knowledge of the \sysnameshortsingle is not required for the adversary to extract profits, and even a limited amount of information can lead to value loss.

\begin{figure}[t!]
    \centering
    \includegraphics[width=\linewidth]{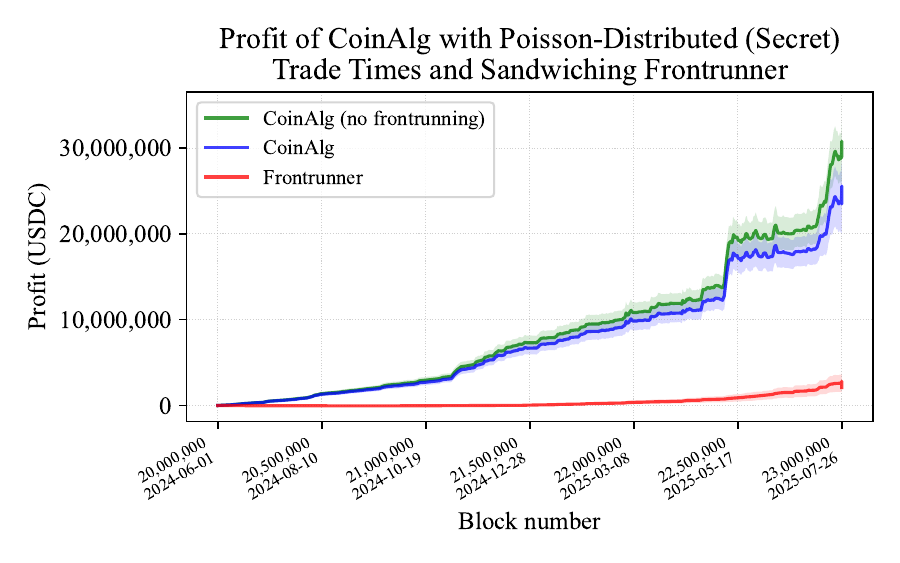}
    \caption{Mean profit across 256 runs for CoinAlg (blue) starting with 300 WETH trading WETH/USDC using randomized, hidden trade times, a frontrunner (red) starting with 200 WETH / 400,000 USDC engaging in long-range sandwich attacks, and the same CoinAlg (green) if a frontrunner were not present. Bands show the range between the 10th and 90th percentiles.}
    \label{fig:fr-profit-poisson}
\end{figure}

We consider this setting by modifying our \sysnameshortsingle so that its trade times are secret: rather than trading on a fixed schedule, it places its trades according to a Poisson process. Unlike traditional sandwich attacks, the frontrunner must now assume risk, since it will hold the \sysnameshortsingle's position for an extended time. We give the frontrunner a simple algorithm: immediately after seeing a trade from the \sysnameshortsingle, it calculates the expected time of the next \sysnameshortsingle trade \iffull{($3{,}600 \text{ blocks}$ in the future) }\fi and attempts to frontrun a trade produced by the \sysnameshortsingle's algorithm for this expected target block. After the \sysnameshortsingle finally trades, it places its backrun transaction. Ultimately, as exemplified in~\Cref{fig:fr-profit-poisson}, the frontrunner still profits substantially.

\begin{figure}[t!]
    \centering
    \includegraphics[width=\linewidth]{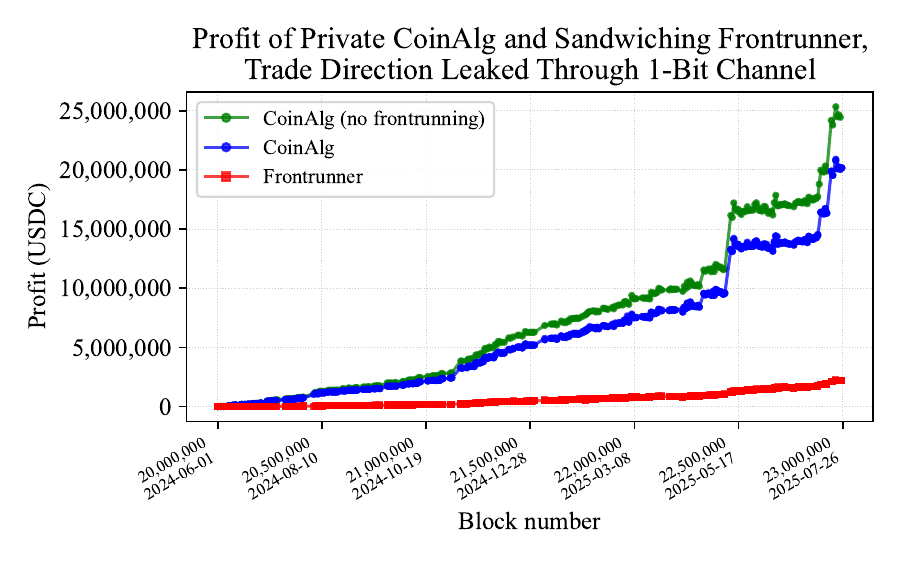}
    \caption{Profit comparison for private CoinAlg (blue) trading WETH/USDC leaking only trade directions to a privileged frontrunner (red) engaging in sandwich attacks estimating 1\% price movements per trade, and the same CoinAlg (green) if there were no frontrunner.}
    \label{fig:fr-profit-covert-1bit-channel}
\end{figure}

We further consider an even more limited adversary, where the \sysnameshortsingle is now private, except for a low-bit covert channel that the adversary has access to. We do not detail the construction of this channel but posit that it may be realized, for example, through an encrypted timing side channel related to the \sysnameshortsingle's processing. Ten blocks prior to its next trading time, the \sysnameshortsingle over this channel communicates to the frontrunner a single bit containing the anticipated direction of its next trade. \iffull{The channel does not communicate the exact trade amount or price limit, or even whether the \sysnameshortsingle will submit a trade, which may not happen if placing a trade would not be profitable (e.g., when the market price is too close to its limit price). }\fi The frontrunner runs a modest sandwiching strategy against the \sysnameshortsingle, blindly expecting the \sysnameshortsingle to make a fixed price impact of 1\%.

As shown in~\Cref{fig:fr-profit-covert-1bit-channel}, the frontrunner remains profitable with this limited information and decreases the \sysnameshortsingle's profits by 17.8\%. \iffull{Though a number of the frontrunner's individual trades were unprofitable, its successful ones more than offset the losses. The impact of this frontrunner on \sysnameshortsingle profits is more than half the impact of a frontrunner on the fully transparent \sysnameshortsingle.}\fi

\paragraph{Collective trades with little market impact are not affected by frontrunning.} In our experiments, when the cost of a trade to the frontrunner exceeded the value which could be extracted from a sandwich trade, the frontrunner opted not to trade. For example, for the 0.05\% USDC/WETH pool, the cost of running a sandwich attack on a liquid pool is at least $0.05\% \times 2 = 0.1\%$ of the input. This cost dominates when available liquidity is high relative to the \sysnameshortsingle's trade size, as also observed in~\cite{zhou2021high}.

\section{Heuristic Guardrails}
\label{sec:guardrails}

Our results in this work show that the \tradeoffName is 
\iffull
inescapable; no mitigation will address it entirely. \sysnameshort (particularly \textit{private} ones) would seem an inevitable part of the investment landscape, however. 
\fi
Thus, finding \emph{guardrails} that limit the impact of the \tradeoffName in private \sysnameshort is crucial. In this section, we sketch a few practical, heuristic guardrails, which we hope motivate future research. 

\paragraph{Randomizing wrappers.} \emph{Randomization} of a \sysnameshortsingle $\DAO$'s trades can make it more challenging for an adversary to extract value from them. In particular, $\DAO$ can select its next trade u\textit{sing private randomness} generated by a \textit{transparent} mechanism that ensures \textit{fairness}, i.e., no leakage to or biasing by an adversary. We call such a mechanism a \textit{randomizing wrapper}. 

\iffull
A randomizing wrapper hybridizes a private \sysnameshortsingle with a transparent guardrail, bringing the profitability benefits of a private trading strategy while limiting opportunities for exploitation through information leakage to an insider.  
\fi

Recall from Section~\ref{sec:definition} that $\DAO$ selects its trade in state $\state$ by running: (1) trade-generation algorithm $\pi_\state \gets \algDist(\state)$, and  (2) sampling algorithm $\trade \getsr \algSample(\pi_\state)$. The idea is to replace (2) and implement a new private \sysnameshortsingle $\tilde{\DAO}$ with these two components: 
\begin{enumerate}
    \item A \textit{private} algorithm $\pi_\state \gets \algDist(\state)$ and
    \item A \textit{transparent} randomizing wrapper $\textsf{RW}$.
\end{enumerate} 

\noindent Here, $\textsf{RW} = (\textsf{RW}_{\$}, \textsf{RW}_e)$, where $\trade \getsr \textsf{RW}_{\$}(\pi_\state)$ selects a transaction using fair, private coins, and $\textsf{RW}_e$ then executes $\trade$. $\textsf{RW}$ is transparent in the sense that its code is published and its properties can be publicly verified. (Its \textit{state}, however, is not transparent during execution.) By running $\tilde{\DAO}$ in a TEE, it is possible to enforce privacy for $\algDist(\state)$ and the distribution of trades $\pi_\state$. At the same time, the TEE can ensure correct execution of $\textsf{RW}$ with fair, secret, TEE-generated random coins. 

\iffull
In summary, $\tilde{\DAO}$ executes a private, randomized trading strategy where $\textsf{RW}$ can be publicly verified by users to ensure that it selects trades using fair randomness that is not leaked to an adversary. 
\else
\fi
While randomizing wrappers cannot completely eliminate attacks, they can help minimize value extraction.

We note that, in some cases, randomizing wrappers can lead the \sysnameshortsingle to perform suboptimal trades, since the wrapper may sample a transaction that is not the most profitable one. However, there \emph{are} cases where randomization does not come at the expense of profits. For example, if a \sysnameshort is performing long-term forecasting, the wrapper can randomize the \emph{time} that trades takes place, effectively acting as a timing jitter. In this case, an adversary would not be able to leverage the instantaneous price increase required for a profitable sandwich attack.

\paragraph{Protected training pipelines.} Randomizing wrappers modify the \emph{runtime} behavior of a \sysnameshortsingle to minimize value extraction. An alternative approach is to modify a \sysnameshortsingle at \emph{creation} time. As a concrete example, if the \emph{data sources} used to train an AI-based \sysnameshortsingle were public, this would provide assurance that the model has not been engineered in malicious ways during the training process. Even if the list of data sources is transparent, however, a key technical challenge is guaranteeing that the model has actually been trained (only) on data from these sources. To achieve this, \sysnameshort can be trained using \emph{protected pipelines}~\cite{juels2024props} which securely attest to the data sources that AI-models have been trained on.

\paragraph{Bug bounty systems.} Complementary to  technical countermeasures is the approach of \emph{disincentivizing} exploitation of the \tradeoffName via the \emph{economic} means of a \emph{bug bounty} system.
The idea is to enable users to prove the existence of value-extraction backdoors and claim monetary rewards, incentivizing close scrutiny of \sysnameshort or even whistleblowing. In the context of \sysnameshortsingle vulnerabilities, such a bug-bounty system can be made publicly verifiable---even as a smart contract~\cite{breidenbach2018enter}. A claimant submits a sequence of correct predictions relating to a \sysnameshortsingle \DAO's trades. With enough prediction entropy, the sequence constitutes a sound proof of access to a covert (or side) channel.

As a simple, concrete example, consider a bounty system $\bountysystem$ for trade-information leakage from a randomizing wrapper $\textsf{RW}$ in the construction above.  Suppose a prover / whistleblower $\player$ claims to learn a predicate $F(\trade)$ prior to $\DAO$'s execution of any trade $\trade$. 
The proof system then works as follows. $\player$ sends $F$ to $\bountysystem$. Then, in each of a sequence $i \in [1,t]$ of epochs, $\player$ sends to $\bountysystem$ a prediction $x_i$ of $F(\trade_i)$ prior to execution of $\trade_i$.
In epoch $t+1$, $\bountysystem$ checks two 
\iffull
conditions:

\begin{enumerate}
    \item \textit{Predictions are correct:} $\forall i \in [1,t]: x_i = F(\trade_i)$.
    \item \textit{Soundness threshold is met:} Prediction sequence $\{x_i\}$ yields a proof with soundness $1-2^{\theta}$ for system parameter $\theta$ (e.g., $\theta=128$).
\end{enumerate}
\fi
\noindent Let $X_i = F\big(\trade_i \getsr \textsf{RW}_{\$}(\pi_{\state_i})\big)$ denote a random variable over the private coins of $\textsf{RW}_{\$}$. To verify condition (2), $\bountysystem$ passes $(F, t, \theta)$ to an API---integrated  into $\textsf{RW}$ to support the bounty system---which checks 
\iffull
that:
\begin{equation*}
\sum_{i=1}^{t}H_\infty(X_i) \geq \theta.
\end{equation*}
\fi

\section{Related Work}
\label{sec:related}

\paragraph{\sysnameshort.} While not termed as such, \sysnameshort in TradFi have been studied extensively, as they encompass various financial instruments (see Section~\ref{sec:what-are-coinalgs}). For example, previous work examines the efficient-market hypothesis (i.e., market-beating performance is impossible using only public information~\cite{timmermann2004efficient,malkiel2005reflections}); and risks from opaque instruments have motivated research on investor protection and fiduciary duties~\cite{laby2010fiduciary, sato2014opacity}, including for robo-advisors~\cite{ji2017robots}. However, no prior work has jointly studied privacy \emph{and} 
\iffull
transparency or their inherent tension.
\fi

Literature on Web3 \sysnameshort is limited, focusing mostly on AI-powered DAOs. Notably, Patlan et al.~\cite{patlan2025ai} conduct a security analysis of ElizaOS and expose prompt-injection vulnerabilities. These results are orthogonal and complementary to ours, as they show different concerns of AI-powered DAOs.

\paragraph{Sandwich attacks.} Sandwich attacks have received substantial prior study. Most relevant to our work, several papers analyze sandwich attacks game-theoretically, as we do in Section~\ref{sec:transparency}. Heimbach and Wattenhofer~\cite{heimbach2022eliminating} introduce a sandwich game to formally analyze such attacks, but, critically, their analysis is one-shot while we consider the infinitely-repeated variant. Zhou et al.~\cite{zhou2021high} similarly formalize sandwich attacks, but limit their analysis to attackers, omitting victim-side considerations such as slippage limit adjustments. 
\iffull
Other works study sandwich attacks empirically, similar to Section~\ref{sec:empirical_transparency}. Many papers~\cite{qin2022quantifying,torres2021frontrunner,zhou2021high} quantify value extracted via sandwiching and other MEV attacks, but our experiments add new market participants rather than merely observing historical data, since private \sysnameshort are not observable and truly transparent \sysnameshort do not yet exist.
\else
\fi

\paragraph{Repeated games.}
\iffull
A core component of the results in Section~\ref{sec:transparency} is the analysis of the \emph{repeated} variant of our sandwiching game.
\else
\fi
Repeated games are surprisingly sparse in DeFi literature, as most works consider one-shot or finite-horizon games (e.g.,~\cite{heimbach2022eliminating,laszka2015bitcoin}). Yet, DeFi involves repeated, strategic interactions between players, and so repeated games are highly relevant. A few works do consider repeated games. For example, Daian et al.~\cite{daian2020flash} model ``priority gas auctions'' with successive adaptive
\iffull 
bidding; while their analysis is one-shot, each iteration resembles a repeated game. Similarly, Judmayer et al.~\cite{judmayer2022estimating} explore MEV definitions using repeated games to model blockchain assets.
\fi

\paragraph{Crypto and AI.} Our paper contributes to the emerging intersection of crypto and AI, such as on ``DePINs'' for AI computations~\cite{lin2024decentralized}, payments for agentic AI~\cite{marino2025giving}, 
\iffull
``trustless''
\else
\fi
agent-to-agent protocols~\cite{erc8004}, authenticated ML pipelines~\cite{juels2024props}, and more. Crypto and  AI are meeting at a brisk pace, and the \tradeoffName is at the core of this convergence, as any AI-based financial instrument involving multiple users could be in the scope of our results.

\section{Conclusion}
\label{sec:conclusion}
This work begins the study of the \tradeoffName, the tradeoff between profitability and economic fairness in \sysnameshort, a broad class of financial instruments. We argue that \sysnameshort can either be transparent, and risk losing profits; or be private, and open the door for unfair value extraction by insiders. Through extensive experiments, we show that this threat is not merely a theoretical concern---even in scenarios where a \sysnameshortsingle may seem benign, subtle backdoors can be inserted that lead to profits from adversaries. Regardless of this fundamental tension, however, \sysnameshort are an inevitable part of the financial landscape, and thus the search for principled guardrails remains an important avenue for future work.

\section*{Acknowledgements}
This work was funded by NSF CNS-2112751 and generous support from IC3 industry partners and sponsors. Ari Juels serves as Chief Scientist at Chainlink Labs.

\bibliographystyle{plain}
\bibliography{references}

\appendix
\section{Proof of Theorem~\ref{thm:unfair_implies_private}}\label{sec:A-unfairness-privacy-proof}

\begin{proof}
    Recall that Definition~\ref{def:privacy} states that, for any set of states $\stateSpaceSubset$, distributions $\pi_\state$ (from which $\DAO$ samples its trade) and $\pi^\leakage_\state$ (the possible trades from $\DAO$ induced by the public function) have average statistical distance at least $\epsilon$. An equivalent way to state this definition is that there exists a \emph{distinguisher} $\distinguisher$ which can distinguish samples from both distributions with probability at least $\epsilon$. That is:
    \begin{align*}
        \Big|&\Prob{x \getsr \pi_\state : \distinguisher(x) = 1} - \\
        &\Prob{x \getsr \pi^\leakage_\state : \distinguisher(x) = 1}\Big| \geq \epsilon
    \end{align*}

    Our goal is thus to show the existence of this distinguisher. Intuitively, $\distinguisher$ will use the fact that, by the definition of economic unfairness, there exists an adversarial algorithm that extracts more value when using the perfect-prediction oracle $\oracle$ (i.e., when receiving samples from $\pi_\state$), than any other algorithm that receives samples from $\pi_\state$ instead.

     By assumption, since there exist $\alpha, t > 0$ such that $\DAO$ is $(\alpha, t)$-unfair with respect to $\oracle$, there exists an algorithm $\advA$ such that, for every possible efficient algorithm $P$, $\advA^\oracle$ extracts $\alpha$ units of additional value than $P^\leakage$ with probability $t$. Fix some $P$, such as the algorithm that achieves the maximum value using $\leakage$. $\distinguisher$ will run the $\FairnessGame$ game with $\advA$ and $P$ as inputs, using its input distribution $\tilde{\pi}$ (which it is trying to distinguish) to respond to queries from both players. Concretely, on every query from $\advA$, $\distinguisher$ will sample trade $x \getsr \tilde{\pi}$ and return $x$ to $\advA$. Note that this is the same distribution as what $\advA$ gets from $\oracle$ if and only if $\tilde{\pi} = \pi_\state$. Then, on every query from $P$, $\distinguisher$ will again sample $x \getsr \tilde{\pi}$, but this time return the property of $x$ that is revealed by public function $\leakage$ to $P$. Note that this is the same distribution as what $P$ gets from its oracle, irrespective of the value of $\tilde{\pi}$. Finally, $\distinguisher$ outputs 1 if, at the end of the game, $\advA$ extracts more than $\alpha$ value. By assumption, $\distinguisher$ outputs 1 when it gets samples from $\pi_\state$ with probability $t$. Conversely, it will output 1 with probability 0 when it gets samples from $\pi^\leakage_\state$. We thus arrive at our desired result.

\end{proof}

\section{Proofs for Results on Transparent \sysnameshort}

\subsection{Proof of Lemma~\ref{lem:cost-buffer}}\label{subsec:A-cost-buffer-proof}

\begin{proof}
The initial AMM reserves are \mbox{$\balAMM_0 = \pairbal{pr_0}{r_0}$}. Lifting the TOK price to \mbox{$p + \plim$} results in new balance \mbox{$\balAMM_1 = \pairbal{(p+\plim)r_1}{r_1}$}. As this is a constant-product AMM, we have:

\begin{equation*}
    (p+\plim)r_1^2 = pr_0^2. 
\end{equation*}

Thus,

\begin{equation*}
    r_1 = r_0\sqrt{\frac{p}{p+\plim}}. 
\end{equation*}

The USD transaction cost is:

\begin{align}
\balAMM_1&(\mathrm{USD}) - \balAMM_0(\mathrm{USD})
  = (p+\plim)\,r_1 - p\,r_0 \notag \\[4pt]
  &= (p+\plim)\,r_0\sqrt{\frac{p}{p+\plim}} - p\,r_0 \notag \\[4pt]
  &= p\,r_0\!\left(\sqrt{\frac{p+\plim}{p}}-1\right).
\end{align}
\end{proof}

\subsection{Proof of Lemma~\ref{lem:cost-buffer-stylized}}\label{subsec:A-cost-buffer-stylized}

\begin{proof}
    \begin{align*}
        p\,r_0\!\left(\sqrt{\frac{2p+e}{p}}-1\right) &\leq p\,r_0 \sqrt{\frac{2p+e}{p}} \\
        &\leq p\,r_0 \left(\sqrt{\frac{2p}{p}} + \sqrt{\frac{e}{p}}\right) \\
        & = p\,r_0 \sqrt{2} + \frac{p}{\sqrt{p}} \sqrt{e} \\
        & = p\,r_0 \sqrt{2} + \sqrt{p} \sqrt{e} \\
        & \leq p\,r_0 \sqrt{2} + p\,r_0 \sqrt{2} \sqrt{e} \\
        & \leq p\,r_0 \sqrt{2} \sqrt{e} + p\,r_0 \sqrt{2} \sqrt{e} \\
        & = z \sqrt{e}\text{, where } z = p\,r_0 2\sqrt{2}
    \end{align*}
\end{proof}

\subsection{Proof of Theorem~\ref{thm:ult-game-eq}}\label{subsec:A-transparency-proof}

\begin{proof}
Let $0 < e^* \leq s$ be some fixed cutoff point. Consider the following strategy profile $\Sigma$:
\begin{enumerate}
  \item \textbf{DAO~$D$:} In every round $t$, divide $s$ as $(e_t, s-e_t)$ for some $e_t = e^*$, provided $b = 1$ in all prior rounds; otherwise set $e_t = \epsilon$ for some minimum value $\epsilon$.
  \item \textbf{Arbitrageur $\advA$:} If round $t$ satisfies
    $e_t = e^*$, output $b= 1$. \emph{Credibly commit} to outputting $b = 0$ for all rounds following $e_t < e^*$.
\end{enumerate}

Let us analyze the payoffs of this strategy profile. Along the cooperative path, the payoff streams for $\DAO$ and $\advA$ are:
\begin{align*}
    \Pi_{\DAO} = \sum_{t=0}^{\infty} c_1(s - e_t) \discount_\DAO^{t} &=  \sum_{t=0}^{\infty} c_1(s - e^*) \discount_\DAO^{i} = \frac{c_1(s - e^*)}{1-\discount_\DAO}\\
    \Pi_{\advA} = \sum_{t=0}^{\infty}  c_2e_t \discount_\advA^{i} &= \sum_{t=0}^{\infty}  c_2e^* \discount_\advA^{i} = \frac{c_2e^*}{1-\discount_\advA} 
\end{align*}

Conversely, let us say that $\DAO$ deviates from the cooperative path in some round $\tilde{t}$, i.e., selects $e_{\tilde{t}} < e^*$. If so, $\DAO$ receives a payoff of 0 in all future rounds, and thus its total payoff is strictly less than in the cooperative path. Importantly, by \emph{committing} to use the grim trigger upon defection, $\advA$ will punish $\DAO$ even though this comes at a cost: without this commitment, the grim trigger would not be a credible threat, and thus it would in fact be rational to accept $\DAO$'s defection. Therefore, $\DAO$ will stay in the cooperative path, as desired.

\end{proof}

\end{document}